\documentclass{ws-ijmpa}
\usepackage[super,compress]{cite}
\usepackage{dcolumn}
\usepackage{bm}
\usepackage{xcolor}

\usepackage[utf8]{inputenc} 
\usepackage[T1]{fontenc}    
\usepackage{hyperref}       
\usepackage{url}            
\usepackage{booktabs}       
\usepackage{nicefrac}       
\usepackage{microtype}      
\usepackage{mathtools}
\usepackage{commath}
\usepackage[sc,osf]{mathpazo}

\usepackage{float}
\usepackage{varioref}
\usepackage[sans]{dsfont}
\usepackage{tikz}

\usepackage[makeroom]{cancel}
\usepackage{graphicx}
\DeclareMathOperator{\sign}{sign}
\newcommand{\Mathematica}{\textit{Mathematica\textsuperscript{\resizebox{!}{0.8ex}{\textregistered}}}}
\usepackage{xspace}
\def\8{\infty}
\def\oh{\frac{1}{2}}
\def\ot{\frac{1}{3}}

\def\tt{\frac{2}{3}}

\def\twfi{\frac{2}{5}}
\def\thfi{\frac{3}{5}}

\def\d{\partial}
\def\i{\imath\,}

\def\dal{\partial_{\alpha}}
\def\dbe{\partial_{\beta}}
\def\dga{\partial_{\gamma}}

\def\undertext#1{\vtop{\hbox{#1}\kern 1pt \hrule}}
\def\ra{\rightarrow}
\def\Ra{\Rightarrow}

\def\VEV#1{\left\langle #1\right\rangle}
\def\VVEV#1{\left\langle\left\langle #1\right\rangle\right\rangle}
\def\tr{\hbox{tr}\,}

\def\pp#1{\frac{\partial}{\partial#1}}
\def\pbyp#1#2{\frac{\partial#1}{\partial#2}}
\def\ff#1{\frac{\delta}{\delta#1}}
\def\fbyf#1#2{\frac{\delta#1}{\delta#2}}

\def\bea{\begin{eqnarray} & &}
\def\eea{\end{eqnarray}}

\def\EXP#1{\exp\left(#1\right)}

\def \R{\mbox{Re}}

\def\CL{Clebsch}

\def\KO{Kolmogorov}
\def\NS{Navier-Stokes}
\def\KH{Kelvin-Helmholtz}
\def\BS{Biot-Savart}

\def\DS {discontinuity surface}

\def\val{v_{\alpha}}
\def\vbe{v_{\beta}}
\def\vga{v_{\gamma}}

\def\ral{r_{\alpha}}

\def\rbe{r_{\beta}}
\def\rga{r_{\gamma}}
\def\rla{r_{\lambda}}
\def\oal{\omega_{\alpha}}
\def\obe{\omega_{\beta}}

\def\XXint#1#2#3{{\setbox0=\hbox{$#1{#2#3}{\int}$}
     \vcenter{\hbox{$#2#3$}}\kern-.5\wd0}}

\DeclareMathOperator{\erf}{erf}

\begin{document}


\title{Vortex Sheet Turbulence as Solvable String Theory}
\author{Alexander Migdal}
\address{Department of Physics, New York University \\
  726 Broadway, New York, NY 10003}
\maketitle
\begin{abstract}
  We study steady vortex sheet solutions of the Navier-Stokes in the limit of vanishing viscosity at fixed energy flow. 
  We refer to this as the turbulent limit.
  These steady flows correspond to a minimum of the Euler Hamiltonian as a functional of the tangent discontinuity of the local velocity parametrized as $\Delta \vec v_t =\vec \nabla \Gamma$. This observation means that the steady flow represents the low-temperature limit of the Gibbs distribution for vortex sheet dynamics with the normal displacement $\delta r_\perp$ of the vortex sheet as a Hamiltonian coordinate and $\Gamma$ as a conjugate momentum.
  An infinite number of Euler conservation laws lead to a degenerate vacuum of this system, which explains the complexity of turbulence statistics and provides the relevant degrees of freedom (random surfaces).
  \textbf{The simplest example of a steady solution of the Navier-Stokes equation in the turbulent limit is a spherical vortex sheet whose flow outside is equivalent to a potential flow past a sphere, while the velocity is constant inside the sphere. Potential flow past other bodies provide other steady solutions. The new ingredient we add is a calculable gap in tangent velocity, leading to anomalous dissipation}.
  This family of steady solutions provides an example of the Euler instanton advocated in our recent work, which is supposed to be responsible for the dissipation of the \NS{} equation in the turbulent limit. We further conclude that one can obtain turbulent statistics from the Gibbs statistics of vortex sheets by adding Lagrange multipliers for the conserved volume inside closed surfaces, the rate of energy pumping, and energy dissipation. The effective temperature in our Gibbs distribution goes to zero as $\mbox{Re}^{-\frac{1}{3}}$ with Reynolds number $\mbox{Re}\sim \nu^{-\frac{6}{5}}$ in the turbulent limit.
  \textbf{The Gibbs statistics in this limit reduces to the solvable string theory in two dimensions (so-called $c=1$ critical matrix model)}. This opens the way for non-perturbative calculations in the Vortex Sheet Turbulence, some of which we report here.
\end{abstract}

\section{\label{sec:level1} Introduction}

Tangent velocity discontinuity has been around for ages.
Make the water above and below a planar surface move fast in the opposite direction, and this discontinuity surface arises and then becomes unstable.

This phenomenon is the famous \KH{} instability\cite{KH20}. The time evolution of this instability leads to the surface rolling up in one direction, creating a vorticity layer.

There were many simulations of this process, starting from the 1800's. Initially, it was analyzed as a 1-dimensional Birkhoff-Rott equation, neglecting one of the coordinates of the planar interface\cite{MBO82}. This equation showed finite-time singularity instead of turbulence, so it was eventually rejected.

It was then realized \cite{M88,AM89} that this was a nonlinear 2D problem of a vorticity sheet moving in a self-generated velocity field. There are no infinities, just a tangent discontinuity of the velocity, so that the Hamiltonian, momentum, and other conserved quantities in that nonlinear mechanical system are finite. The Action principle was found in \cite{M88,AM89}, reproducing Lagrange dynamics in this particular case, with an infinite number of conservation laws related to the Kelvin theorem of conservation of circulation.

Later, there were several such studies\cite{Kaneda90, Kuz18}, with more references inside. The verdict was the same -- this motion is unstable. It leads to a vorticity layer, ending with turbulence, which remains unsolved after centuries of good tries.

However, to our knowledge, nobody noticed that this surface discontinuity in Euler dynamics could be steady under certain conditions. This steady Euler flow is the primary subject of the present paper. 

By steady we do not mean stable, on the contrary, we believe in the \KH{} instability as a general mechanism of the turbulence onset. We expect the vortex sheet as a dynamical system to go around the energy surface in its phase space and eventually cover it with some invariant measure, just because there are no stable attractors in this space.  In the zeroth approximation, we expect the ergodicity of the vortex sheet dynamics, uniformly covering the energy surface in phase space.

The viscous dissipation of energy leads to the following modification of this ergodic dynamics. We have to intersect this energy surface with two other  surfaces: one for the fixed energy pumping and another for equal fixed energy dissipation. These notions were introduced in \cite{M20c} and will also be explained in this paper from a slightly different point of view. With these two extra constraints, the Euler statistics becomes equivalent to a certain Gibbs distribution, which we then reduce to a solvable string theory.

There is a long path from the \KH{} instability of a vortex sheet to the end statement, which is a very specific solvable string theory which we find. 

First, we study the Euler-Lagrange dynamics \cite{M88,AM89} of the vortex sheets (in this paper we assume these sheets to be closed surfaces). We find a very simple method of computing steady Euler solutions for vorticity and velocity fields parametrized by an arbitrary set of surfaces.

This infinite ambiguity of steady solutions is the result of the initial ambiguity of an infinite number of conservation laws in the Euler-Lagrange dynamics. The vortex structures (circulation around fluid loops, to be more precise) are conserved, being passively moved, frozen in the flow.

In the vortex sheet dynamics, this corresponds to the density $\Gamma = \Phi^+ - \Phi^-$  where $\Phi^\pm$ are the velocity potentials at two sides of the surface. The velocity field outside the vortex surface is purely potential $\vec v = \vec \nabla \Phi$ because there is no vorticity outside.

These potentials satisfy the Laplace equations with Neumann boundary conditions, which we use to find some steady flows.
The density $\Gamma$ needed for the steady Euler solution, satisfies a certain linear integral equation, corresponding to the minimization of the Hamiltonian. Thus, we trade the ambiguity of $\Gamma$ for the ambiguity of the shapes of these surfaces.

The steady solutions of the Euler equations hide the fact that in the Lagrange dynamics these surfaces move with constant speed without changing their shape, but with time-dependent diffeomorphism of internal coordinates.  Therefore, this is a steady Lagrange solution up to a Galilean transformation and time-dependent diffeomorphism.

These are very basic statements about the Euler-Lagrange dynamics of vortex sheets, not using any of the modern developments of the theory of turbulence. These mechanical statements could have been established 200 years ago, but were overlooked by our great predecessors.

We argue that this is also a solution of the \NS{} equation in a turbulent limit, as in this limit the thickness of the vortex sheet goes to zero and the singular viscous terms in the \NS{} equation are exactly cancelled by the singularities of the Euler terms.

In this paper, we go one step further and find all terms (singular and regular) in the solution of the \NS{} equation in the tangent frame of coordinates, with $z$ directed at the surface normal. 

This solution is an old Burgers vortex sheet\cite{BM05}, rotated and translated in the tangent plane, to match the Euler solution with a tangent discontinuity at $\frac{z}{h} \ra \pm \infty$. The vorticity has a Gaussian profile as a function of the normal coordinate $z$, with some universal width $h$ going to zero as some power of viscosity in a turbulent limit.

For the sphere, we find explicit formulas and study the correspondence between the Lagrange and Euler solutions and derive the explicit form of the time-dependent diffeomorphism (the Cauchy problem solution for the Lagrange dynamics). 

For our final goal of statistical distribution, such quasi-steady solutions are sufficient.

We discuss the topological properties of vortex sheets, such as their self-avoidance and the topological nature of velocity circulation around the handle of the sheet such as a torus.

Then, we revisit the viscosity anomaly (persistence of dissipation in the turbulent limit) in the context of this vortex dynamics, and we rederive the scaling laws in the turbulent limit of viscosity going to zero at a fixed energy flow. 

For the reader's convenience, we derive this anomaly from scratch, as this novel notion is not yet widely accepted. Besides, this solution is now modified using the Burgers exact solution of the \NS{} equation in a local tangent frame to the surface, up to the small curvature effect at vanishing width $h$. We match all terms in the local \NS{} equation, not just the singular ones.

The implication of these scaling laws is that the turbulence arises spontaneously, with the external force needed to provide the energy flow vanishing as some positive power of viscosity. 

Additionally, the effective thickness of the vortex sheet vanishes as another positive power of viscosity, which justifies the Euler dynamics, but with an anomalous dissipation term in effective Hamiltonian.

Finally, we study the Gibbs-like distribution in our system, which we see as an ensemble of closed vortex surfaces ("bubbles") moving around in space (avoiding each other and themselves) and exchanging viscosity fluxes $\Gamma$ with each other due to small viscous effects. 

These effects also lead to the energy dissipation, so we introduce a random Gaussian uniform force pumping the energy through the boundary at infinity.

There is a nice surprise, opening the way to a quantitative theory-- the effective temperature of this Gibbs ensemble comes out small. It goes to zero as $\R{}^{-\ot}$ with growing Reynolds number.

This energy is dissipated inside the vortex sheets due to the viscosity anomaly, which leads to the emergence of a new degree of freedom, hidden so far, the waves of the vorticity density $\Gamma$, defined as a gapless 2D field on a vortex sheet.

This field interacts with another hidden field-- the Liouville field $\varphi$ coming from an internal metric on the surface. This field emerges in string theory as a result of the conformal invariance of the measure for random surfaces. This conformal invariance, in turn is a remnant of the full diffeomoprphism invariance  remaining after the conformal metric is chosen as a gauge condition.

The fluctuations of the shapes of the surfaces in 3D space are not gapless -- there are some forces coming from surface tension and volume conservation. As a result of these forces, the shape is rather rigid, like a true soap bubble which slowly fluctuates around its spherical form, balancing the inner pressure and surface tension.

However, just as there are those rainbow images flowing effortlessly around the soap bubble, there are two soft modes $\Gamma$ and $\varphi$ living on the vortex surface and strongly fluctuating in the turbulent limit (low temperature).

The final piece of luck: this particular string theory with two components of string $\Gamma,\varphi$ is exactly solvable. It was solved back in the early 90's by means of the Liouville theory and matrix models of gravity.
As a result, we can compute something beyond perturbation theory in this new Vortex Sheet Turbulence.

Let us define here the necessary equations.

\NS{} equations 
\begin{subequations}
\begin{eqnarray}\label{NSv}
   && \d_t \vec v + (\vec v \cdot \vec \nabla) \vec v + \vec \nabla p =  \nu \vec \nabla^2\vec v;\\
   && \vec \nabla \cdot \vec v =0;
\end{eqnarray}
\end{subequations}
can be rewritten as the equation for vorticity
\begin{subequations}
\begin{eqnarray}
&&\vec \omega = \vec \nabla \times \vec v ;\\
    &&\d_t \vec \omega + (\vec v \cdot \vec \nabla) \vec \omega - (\vec \omega \cdot \vec \nabla) \vec v =  \nu \vec \nabla^2\vec \omega;\label{NSE}
\end{eqnarray}
\end{subequations}

As for the velocity, it is given by a Biot-Savart integral
\begin{equation}\label{BS}
    \vec v(r)= -\vec \nabla \times \int d^3 r' \frac{\vec \omega(r')}{4\pi | r - r'|}
\end{equation}
which is a linear functional of the instant value of vorticity.

The energy dissipation is related to the enstrophy (up to total derivative terms, vanishing by the Stokes theorem)
\begin{eqnarray}\label{Diss}
    -\d_t H = \mathcal{E} = \nu \int d^3 r \vec \omega^2 
\end{eqnarray}

One comment about notations. We are going to use the Einstein convention of summation over repeated indexes, accepted in theoretical physics for the last 100 years but not so popular in the mathematical literature.
The Greek indexes $\alpha,\beta,\dots $ with run from $1$ to $3$ and correspond to physical space $R_3$ and the lower case Latin indexes $a,b,\dots $ will take values $1,2$ and correspond to the internal parameters on the surface. 

We shall also use the Kronecker delta $\delta_{\alpha\beta}, \delta_{a b}$ in three and two dimensions as well as the antisymmetric tensors $e_{\alpha\beta\gamma}, e_{a b}$ normalized to $e_{1 2 3} = e_{1 2} =1$.

The Euler equation corresponds to setting $\nu =0$ in \eqref{NSv}, \eqref{NSE}. This limit is known not to be smooth, leading to a statistical distribution of vortex structures, which is the whole turbulence problem. 

We addressed this problem in \cite{M20c}. Our initial goal here was not so ambitious: to study the vorticity sheet dynamics in the Euler-Lagrange equations and the steady vortex sheets that seem impossible. 

As an unexpected by-product of this study, we find the general relation between Gibbs statistics and the turbulent statistics for this vorticity sheet system, which moves us closer to our big goal: the Statistical Theory of turbulence.

\section{Steady \DS{} in the Lagrange Dynamics}

As it is well known, the Euler equation is equivalent to the motion of every point in the fluid with local velocity:
\begin{equation}\label{Lagrange}
    \d_t \vec r = \vec v(\vec r).
\end{equation}
This also applies to every point $\vec r = \vec X(\xi_1,\xi_2)$ on the \DS{}, with the principal value prescription, which was discussed in\cite{M88,AM89}. In other words, the velocity at the surface 
\begin{eqnarray}
    \vec v^S(\vec r) = \oh \left(\vec v\left(\vec X^+(\xi)\right) + \vec v\left(\vec X^-(\xi)\right)\right)
\end{eqnarray}

The normal component $v_n$ of velocity describes the surface's genuine change– its global motion or the change of its shape. In a steady flow, this normal component must vanish at every point of the surface.

The remaining two tangent components $\vec v_t$ of the velocity, on the other hand, just move points along the surface without changing its position and shape. To see that, we rewrite the tangent part $\vec v^S_t $ of the surface velocity $\vec v^S(\vec r)$ as time-dependent re-parametrization $\xi \Ra \xi(t)$:
\begin{eqnarray}\label{ParInv}
    &&\vec v^S_t  = \d_t \vec X(\xi(t)) = \d_a \vec X \d_t\xi^a ;\\
    &&\d_t\xi^a =  g^{a b} \d_b \vec X \cdot \vec v^S_t;\\
    && g_{a b} = \d_a \vec X \cdot \d_b \vec  X;
\end{eqnarray}
Here $g_{a b} $ is an induced metric, and $g^{a b}$ is its inverse. To verify this identity, one has to expand $\vec v^S_t $ in the two local plane vectors $\d_1 \vec X, \d_2 \vec X$ and use two identities 
\begin{eqnarray}
    &&\d_b \vec X \cdot \d_c \vec X = g_{b c};\\
    && g^{a b} g_{b c} = \delta_{a c}
\end{eqnarray}

For a closed surface, these tangent motions will never leave the surface. For the surface with the fixed edge $C$ there is a boundary condition that velocity normal to the edge vanishes $ \vec v_t\times d \vec r =0; \forall \vec r \in C$.  In this case, the fluid will slide along $C$ leading to its re-parametrization, but it never leaves the surface.

We will restrict ourselves to the parametric invariant functionals, which do not depend on this tangent flow, and therefore, they will stay steady. 

We introduced the Lagrange action of vortex sheets in the old paper\cite{M88}. In that paper, we also conjectured the relation of turbulence to Random Surfaces.\footnote{I could not find \cite{M88} anywhere online, but it exists in university libraries such as Princeton University Library and U.C. Berkeley Library.}

In the next paper\cite{AM89} this Lagrange vortex dynamics was simulated using a triangulated surface. We calculated the contributions to the velocity field from each triangle in terms of elliptic integrals.  The positions of the triangle vertices served as dynamical degrees of freedom. 

There were conserved variables related to the velocity gap as a function of a point at the surface. These points were passively moving together with the surface by the mean value of velocity on the surface's two sides. 

Later, our equations were recognized and reiterated in traditional terms of fluid dynamics\cite{Kaneda90} and simulated with a larger number of triangles \cite{Brady98} with similar results regarding the \KH{} instability of the vortex sheet. There were dozens of publications using various versions of the discretization of the surface and various simulation methods.

Let us reproduce this theory here for the reader's convenience before advancing it further. 

The following ansatz describes the vortex sheet vorticity: 
\begin{equation}
   \vec \omega(\vec r) = \int_\Sigma d\vec \Omega \delta^3\left(\vec X-\vec r\right)
\end{equation}
where the 2-form
\begin{eqnarray}
    d \vec \Omega \equiv d \Gamma\wedge d\vec X = d\xi_1 d \xi_2 e_{a b} \pbyp{\Gamma}{\xi_a} \pbyp{\vec X}{\xi_b} ;
\end{eqnarray}

This vorticity is zero everywhere in space, except the surface, where it is infinite. To describe the physical vorticity of the fluid, this ansatz must satisfy the divergence equation (the conservation of the "current" $\vec \omega$ in the language of statistical field theory)
\begin{eqnarray}
    \vec \nabla \cdot \vec \omega =0;
\end{eqnarray}
This relation is built into this ansatz for arbitrary $\Gamma(\xi)$, as can be verified by direct calculation. In virtue of the singular behavior of the Dirac delta function, it may be easier to understand this calculation in Fourier space
\begin{eqnarray}
&& \vec \omega^F(\vec k) = \int d^3 r e^{\i \vec k \cdot \vec r} \vec \omega(\vec r) = \int_\Sigma d \vec \Omega e^{\i \vec k \cdot \vec X};\\
    &&\i \vec k \cdot \vec\omega^F(\vec k) =\int_\Sigma  d \Gamma \wedge d \vec X \cdot (\i \vec k) e^{\i \vec k \cdot \vec X}=
     \int_\Sigma d \Gamma\wedge d e^{\i \vec k \cdot \vec X} = \int_{\d \Sigma} d\Gamma e^{\i \vec k \cdot \vec X};
\end{eqnarray}
In case there is a boundary of the surface, this $\Gamma(\xi)$ must be a constant at the boundary for the identity $\vec k \cdot  \omega^F(\vec k)=0$ to hold. Transforming back to $R_3$ from Fourier space we confirm the desired relation $\vec \nabla \cdot \vec \omega =0$ in a sense of distribution, like the delta function itself.

It may be instructive to write down an explicit formula for the tangent components of vorticity in the local frame, where $x,y$ is a local tangent plane and $z$ is a normal direction
\begin{eqnarray}
   && \omega_j(x,y,z) =  \d_i\Gamma  e_{i j}  \delta(z);\\
   && \omega_z(x,y,z) =0;
\end{eqnarray}
In particular, outside the surface, $\vec \omega =0$, so that its divergence vanishes trivially.

The divergence is manifestly zero in this coordinate frame 
\begin{eqnarray}
    \vec \nabla \cdot \vec \omega = \delta(z)\d_j \d_i\Gamma  e_{i j} =0;
\end{eqnarray}
Let us compare this with the \CL{} representation
\begin{equation}
    \oal =  e_{\alpha\beta\gamma} \dbe \phi_1 \dga\phi_2,
\end{equation}
We see that that in case $\phi_2$ takes one space-independent value $\phi_2^{in}$ inside the surface and another space-independent value $\phi_2^{out}$ outside, the vorticity will have the same form, with 
\begin{eqnarray}
    \Gamma = \phi_1 (\phi_2^{in} - \phi_2^{out})
\end{eqnarray}

Neither \cite{AM89} nor any subsequent papers noticed this relation between \CL{} field discontinuity and vortex sheets.

In this paper, except the last section, we study the vortex sheets by themselves as a generalized Hamiltonian system without using the Clebsch variables.

As we already noted in \cite{M88,AM89}, the function $\Gamma(\xi_1, \xi_2)$ is defined modulo diffeomorphisms $\xi \Ra \eta(\xi); \det {\d_i \eta_j} >0$ and is conserved in Lagrange dynamics:
\begin{equation}
    \d_t \Gamma =0;
\end{equation}
This function is related to $1$-form of velocity discontinuity
\begin{equation}\label{DiscGamma}
    d \Gamma  = \Delta \vec v \cdot  d \vec X
\end{equation}
where the velocity gap
\begin{equation}
    \Delta \vec v(\xi) = \vec v\left(X_+(\xi)\right) -\vec v\left(X_-(\xi)\right)
\end{equation}

Another way of writing the relation for $\Gamma$ is to note that the velocity field is purely potential outside the surface, although the potential $\Phi^-$ inside is different from the potential $\Phi^+ $ outside. This is a direct consequence of the fact that vorticity is zero everywhere except at the surface.

In this case, the velocity discontinuity equals the difference between the gradients of these two potentials, or, equivalently
\begin{eqnarray}
    \Gamma(\vec r) = \Phi^+(\vec r) - \Phi^-(\vec r); \forall \vec r \in S
\end{eqnarray}

The surface is driven by the self-generated velocity field (mean of velocity above and below the surface), as in \eqref{Lagrange}. Let us substitute our ansatz for vorticity  into the Biot-Savart integral for the velocity field  and change the order of integration
\begin{eqnarray}\label{BSGamma}
    &&\vec v(\vec r) = -\frac{1}{4 \pi}  \vec \nabla \times\int d^3 r' \frac{1}{|\vec r - \vec r'|}\int d \vec \Omega   \delta^3(X - \vec r')= \nonumber \\
    && \frac{1}{4 \pi}\int d \vec \Omega \times \vec \nabla \frac{1}{|\vec r - \vec X|}
\end{eqnarray}

The Lagrange equations of motion for the surface
\begin{equation}
    \d_t \vec X(\xi) = \vec v\left(\vec X(\xi)\right)
\end{equation}
were shown in\cite{M88,AM89} to follow from the action
\begin{eqnarray}\label{action}
   && S = \int \Gamma d  V- \int H d t;\\
   && d V = d \xi_1 d \xi_2 d t  \pbyp{\vec X}{\xi_1} \times \pbyp{\vec X}{\xi_2} \cdot \d_t \vec X ;\\
   && H = \oh \int d^3 r \vec v^2 = \frac{1}{2} \int_{S}\int_{S}  \frac{ d \vec \Omega\cdot d \vec \Omega' }{4 \pi |\vec X - \vec X'|}; 
\end{eqnarray}
This $d V$ is the 3-volume  swept by the surface area element in its movement for the time $ d t$.

The easiest way to derive the vortex sheet representation for the Hamiltonian is to go in Fourier space where the convolution becomes just a multiplication and use the incompressibility condition $\vec k \cdot \vec v^F(\vec k)=0$
\begin{eqnarray}
    && \vec \omega^F(\vec k)  = \i \vec k \times \vec v^F(\vec k);\\
    &&\vec v^F(\vec k) \cdot \vec v^F(-\vec k) = \frac{\vec \omega^F(\vec k) \cdot \vec \omega^F(-\vec k)}{\vec k^2}
\end{eqnarray}
In the case of the handle on a surface, $\Gamma$ acquires extra term $\Delta \Gamma = \oint_{\gamma} \Delta \vec v \cdot d \vec r$ when the point goes around one of the cycles $\gamma = \{\alpha, \beta\}$ of the handle. 

This $\Delta \Gamma$ does not depend on the path shape because there is no normal vorticity at the surface, and thus there is no flux through the surface. This topologically invariant $\Delta \Gamma$ represents the flux through the handle cross-section.

This ambiguity in $\Gamma$ makes our action multivalued as well.

Let us check the equations of motion emerging from the variation of the surface at fixed $\Gamma$:
\begin{eqnarray}
    &&\delta \int H d t = \int d \vec \Omega \times \delta \vec X\cdot \vec v(\vec X) d t;\\\label{deltaH}
    &&\delta \int \Gamma d V = \int d \vec \Omega \times \delta \vec X\cdot \d_t \vec X d t\label{deltaS}
\end{eqnarray}

As we already discussed above, the tangent components of velocity at the surface create tangent motion, resulting in the surface's re-parametrization.

One of the two tangent components of the velocity (along the line of constant $\Gamma(\xi)$) does not contribute to variation of the action, so that the correct Lagrange equation of motion following from our action reads
\begin{equation}
    \d_t \vec X(\xi) = \vec v(\vec X(\xi)) \mod{e^{i j} \d_i\Gamma  \d_j \vec X}
\end{equation}

We noticed this gauge invariance before in\cite{M88}, but now we see that both tangent components of the velocity could be absorbed into the re-parametrization of a surface and therefore do not represent an observable change.

However, the normal component of the velocity must vanish in a steady solution, and this provides a linear integral equation for the conserved function $\Gamma(\xi)$. \footnote{To be more precise, in the case of a uniform global motion of a fluid, the normal component of the velocity field must coincide with the normal component of this uniform global velocity $\vec v^G$. We shall need that global velocity later.}

In the general case, when there is an ensemble of such surfaces $S_n, n =1,\dots  N$ each has its discontinuity function $\Gamma_n(\xi)$. At each point  on each surface $\vec r \in  S_n$, the net normal velocity adding up from  all surfaces, including this one in the \BS{} integral, must vanish:
\begin{eqnarray}
   && \vec \Sigma_n(\xi) \cdot \vec v(\vec X_n(\xi))=0;\\
   && \vec \Sigma_n(\xi) = e^{ i j}\d_i \vec X_n \times \d_j\vec X_n ;\\
   && \vec v(\vec r)  = \frac{1}{4 \pi} \sum_m\int d \vec \Omega_m \times \vec \nabla\frac{1}{|\vec r - \vec X_m|}
\end{eqnarray}
 
 This requirement provides a linear set of $N$ linear integral equations (called Master Equation in \cite{M20c}) relating $N$ independent surface functions $\Gamma_1\dots\Gamma_n$.
 With this set of equations satisfied, the collection of surfaces $S_1 \dots S_N$ will remain steady up to re-parametrization.
 
Here is an essential new observation we are reporting in this paper. 

\textbf{The Master Equation is equivalent to the minimization of our Hamiltonian by $\Gamma_n, n=1\dots N$}
 \begin{eqnarray}\label{Ham}
   &&H[\Gamma,\vec X]= \frac{1}{2}\sum_{n,m} \int_{S_n}\int_{S_m}  \frac{ d \vec \Omega_n\cdot d \vec \Omega_m }{4 \pi |\vec X_n - \vec X_m|}; \\
   && \fbyf{H[\Gamma,\vec X]}{\Gamma_n(\xi)} = \vec \Sigma_n(\xi) \cdot \vec v\left(\vec X_n(\xi)\right);
\end{eqnarray}

The tangential components of velocity $\vec v_t$ are included in the parametric transformations, as noted above. They are equivalent to variations of the Hamiltonian by the parametrization of $\Gamma, \vec X$ and are therefore the tangent components of \eqref{Lagrange}  are satisfied in virtue of parametric invariance. 

Therefore, the normal velocity of the surface in the general case is equal to the Hamiltonian variation by $\Gamma$, as if $\Gamma$ is the conjugate momentum corresponding to the surface's normal displacement. To be more precise, $\Gamma$ in our action \eqref{action} is a conjugate momentum to the volume, which is locally equivalent -- the variation of volume equals the area element times the normal displacement. 

In other words, we can consider an extended dynamical system with the same Hamiltonian \eqref{Ham} but a wider phase space $\Gamma,\vec X \mod {Diff}$. We can introduce an extended Hamiltonian dynamics with our action \eqref{action}. 

This system is degenerate in the sense that for an arbitrary evolution of $\vec X$ providing an extremum of the action, the evolution for $\Gamma$ is absent, i.e., $\Gamma$ is constant. It is a conserved momentum in our Hamiltonian dynamics with volume as coordinate.

This conservation of $\Gamma$ is a consequence of Kelvin's theorem. To see this relation\cite{AM89}, we rewrite this $\Gamma$ as a circulation over the loop $C$ puncturing the surface in two points $A,B$ and going along some curve $\gamma_{A B}$ on one side, then back on the same curve $\gamma_{B A}$ on another side. The circulation does not depend upon the shape of $\gamma_{ A B}$ because there is no normal vorticity at the surface.

Another way to arrive at the conservation of $\Gamma$ is to notice that it is related to the \CL{} field on the \DS{}, as we mentioned in the introduction.

The steady solution for $\vec X \mod {Diff}$ corresponds to the Hamiltonian minimum as a (quadratic) functional of $\Gamma$. 

\section{Does Steady Surface mean Steady Flow?}

There is a subtle difference between the steady \DS{} and steady flow. After all, the flow around a steady object does not have to be steady? There could be time-dependent motions in the bulk of the flow, while the normal component of the flow vanishes at the solid surface (as it always does).

This logic applies to the generic flow around steady solid objects, but it does not apply here. The big difference is that by our assumption, there is no vorticity outside these discontinuity surfaces. 

The  Biot-Savart integral for the velocity field \eqref{BSGamma} is manifestly parametric invariant, if we transform \textbf{both} $\Gamma, \vec X$
\begin{eqnarray}
    &&\Gamma(\xi) \Ra \Gamma(\eta(t,\xi));\\
    && \vec X(\xi) \Ra \vec X(\eta(t,\xi));\\
    && \d_t \eta_a = \phi_a(\eta);
\end{eqnarray}

This transformation describes the flux of coordinates $\eta$ in parametric space with the velocity field $\phi_a(\eta)$.
The tangent flow around the surface is equivalent to such a transformation of $\vec X$, as we demonstrated in \eqref{ParInv}. 

However, in the Lagrange dynamics of vortex sheets, the function $\Gamma(\xi)$ remains constant, not the constant up to re-parametrization, but an absolute constant so that $\d_t \Gamma(\xi)=0$ where time derivative goes at fixed $\xi$.

Therefore, in general, the velocity field $\vec v(\vec r)$ does change when the surface gets re-parametrized, but $\Gamma$ does not. Naturally, one could not get the steady solution without solving some equations first :).

However, in our steady manifold, $\Gamma^*(\xi)$ is related to the surface by the master equation. Let us write it down once again
\begin{eqnarray}
    &&0=\vec \Sigma_n \cdot \sum_m \int_{S_m} d \Gamma^*_m \wedge d \vec X_m \times \vec \nabla_n \frac{1}{|\vec X_n - \vec X_m|} ;\\
    && \vec \Sigma_n = e^{a b} \d_a \vec X_n \times \d_b \vec X_n
\end{eqnarray}
This equation is invariant by the parametric transformation of both variables $\Gamma^*, \vec X$. This means that the solution of this equation for $\Gamma^*$ would come out as a parametric invariant functional of $\vec X$, in addition invariant by translations of $ \vec X$.

As we have seen, this master equation
leads to a vanishing normal velocity $\vec \Sigma_n\cdot \d_t \vec X_n=0$. The remaining tangent velocity leaves $\vec X$ steady up to re-parametrization. Therefore, in virtue of this master equation, the velocity field will also be steady.

We introduced a family of steady solutions of Lagrange equations, parametrized by an arbitrary set of discontinuity surfaces $\vec X_n(\xi)$ with discontinuities $\Gamma^*_n(\xi)$ determined by the minimization of the Hamiltonian.

The surfaces are steady up to time-dependent reparametrization (diffeomorphism). The equivalence of Lagrange and Euler dynamics suggests that these are steady solutions of the Euler equations.

The above arguments are perhaps too formal to accept our steady solution of the Euler equation.
These are weak solutions with tangent discontinuities of velocity, and thus they require some care to investigate the equivalence between the Lagrange and Euler solutions.

In the following section we work out all details for a particular case of a spherical surface.

\section{Conservation Laws and Minimization Problem}
As it is well known, the Euler dynamics has infinitely many integrals of motion. In the dynamics of vortex sheets, these integrals are generated by two-dimensional conserved function $\Gamma_n(\xi)$ on each \DS{} $S_n$.

One can also write down explicit integrals of motion, involving both $\Gamma$ and $X$ variables.
In addition to the Hamiltonian \eqref{Ham}, there is a helicity $\mathcal{H}$, momentum $\vec P$ and  angular momentum $\vec M$
\begin{eqnarray}
   && \mathcal{H} = \int d^3 r  \vec \omega \cdot \vec v = \nonumber\\
   &&\sum_{n,m}\int_{S_n}\int_{S_m} 
    d \vec \Omega_n \cdot d \vec \Omega_m \times \vec \nabla_m\frac{1}{4 \pi |X_n - X_m|};\label{Hel}\\
    && \vec P = \int d^3 \vec r \vec v = \ot \sum_n\int_{S_n} d \vec \Omega_n \times \vec X_n ;\\\label{PGam}
    && \vec M = \int d^3 \vec r \vec v  \times \vec r= \oh \sum_n\int_{S_n} d \vec \Omega_n \vec X_n^2 
\end{eqnarray}
The easiest way to derive these relations is to use Fourier representation for the velocity and expand the exponential in $\vec k $ in both sides:
\begin{eqnarray}
    \int d^3 r \vec v(r) e^{ \i \vec k \cdot \vec r} =  \frac{\i \vec k}{\vec k^2} \times \sum_n \int_{S_n} d \vec \Omega e^{\i \vec k \cdot \vec X}
\end{eqnarray}

For the closed surfaces there is also a conserved volume inside each of them
\begin{equation}\label{Vol}
    V_n = \int_{B: \d B = S_n} d^3 r  = \frac{1}{3} \int_{S_n} d^2 \xi \vec X \cdot e_{i j}\d_i \vec X \times \d_j \vec X
\end{equation}
This volume only depends on the surface, but not on the vorticity density $\Gamma$.

The viscosity anomaly \cite{M20c} coming from resolving $0\times \infty$ in the enstrophy integral \eqref{Diss} breaks the Hamiltonian conservation  
\begin{eqnarray}\label{enstrophy}
   &&\nu \int d^3 r \vec \omega^2  \ra E[\Gamma,\vec X];\\
   && E[\Gamma, \vec X] = \Lambda \sum_n \int_{S_n} \sqrt{g} g^{i j}   \d_i \Gamma_n\d_j \Gamma_n ;\label{EGam}
\end{eqnarray}
The parameter $\Lambda$ was computed in \cite{M20c} by taking limit $\nu \ra 0$ in the \NS{} equation. We found Gaussian profile of vorticity with viscous width $h$ in direction $z$ normal to the surface. 
\begin{equation}
    \delta_h(z) = \frac{1}{h \sqrt{2\pi}}\EXP{- \frac{z^2}{2 h^2}}
\end{equation}

At $\nu \ra 0$ this Gaussian profile is becomes $\delta(z)$, making it equivalent to our representation of vorticity. At finite but small viscosity, the resulting integral over the viscous layer provides an extra factor $\Lambda/\nu$ with
\begin{eqnarray}
    \Lambda = \frac{\nu}{ 2 h\sqrt{\pi}}
\end{eqnarray}
We sketch this calculation below for the reader's convenience.

Few words about the terminology.
In quantum field theory, the term "anomaly" has a special meaning: this is an explicit finite expression for the time derivative of the fluctuating quantity which would be conserved in the absence of fluctuations at infinitesimal scales.

This is exactly what we have here: the time derivative of Euler Hamiltonian is zero unless you take into consideration the viscous scales, which are infinitesimal from the Euler point of view.

The quantum anomalies appear in the axial current in the Standard Model and the gravitational anomaly (finite trace of the stress-energy tensor, violating the conformal invariance of the classical theory).

The quantum anomalies are the space integrals of some local expressions, and so is our viscosity anomaly. It is the surface integral of the square of tangent velocity discontinuity $\Delta \vec v_t = \vec\nabla_t \Gamma$. 

Just as the gravitational anomaly reveals a hidden dynamical variable\cite{Pol81} -- 2D Liouville field $\varphi$ , our viscosity anomaly reveals the hidden fluctuating 2D field $\Gamma$ which is conserved in Kelvin-Helmholtz dynamics\footnote{Analogy goes even further, as this field $\Gamma$ becomes the string coordinate in the 2D string theory with Liouville field being another one (see below)}.

In the steady state of turbulence, all dissipated energy is compensated by the work made by external random forces, which we introduce as a constant Gaussian random vector imposed at infinity as the boundary condition for pressure $p \ra - \vec f \cdot \vec r$.
\begin{equation}
\mathcal{E} = \VEV{\vec f \vec P}
\end{equation}
We have to minimize the free energy
\begin{eqnarray}\label{FreeEnergy}
    F[\Gamma,\vec X,\vec \lambda] = H[\Gamma,\vec X]- \vec \lambda \cdot (\vec P[\Gamma,\vec X]-\vec Q(\vec f));
\end{eqnarray}
with Lagrange multiplier $\vec \lambda$ to be determined from the fixed momentum
\begin{eqnarray}
     \vec P[\Gamma,\vec X] = \vec Q(\vec f);
\end{eqnarray}
Instead of fixing the momentum, we could declare this Lagrange multiplier $\vec \lambda = \vec f$ an external force. This force in our ensemble becomes Gaussian random variable, uniform in space and time-independent. Averaging over Gaussian distribution  of $\vec f$ in our ensemble plays the same role as averaging over an external delta-correlated Gaussian random force in time-dependent Navier-Stokes equations.

We argued in\cite{M20c} that the momentum could be expanded in terms of the external random force $\vec Q(\vec f) = \hat Q\cdot \vec f$ with some constant $3\times 3$ symmetric matrix $\hat Q$ depending on the distribution of vorticity. In this paper, we compute this momentum analytically and prove its linear dependence of $\vec f$.

The best way to solve this minimization problem is to go back to 3D space and use the fact that there is no vorticity in the space except for the discontinuity surfaces.

Let us assume that inside each surface, the velocity field is just a constant vector.
\begin{eqnarray}
    \vec v( \vec r \mbox{ inside } S_n) = \vec C_n;
\end{eqnarray}
This constant velocity is a trivial solution of the Euler equation. \footnote{This is not the most general potential solution, of course. One can add higher polynomials, such as $ B_{\mu\nu} r_\nu$ with some constant symmetric traceless tensor $B_{\mu\nu}$. We restricted ourselves to a constant velocity to give an example of the steady Euler flow with energy pumping and dissipation. 
In the general case, the inside velocity would be a polynomial of $\vec r$, restricted by the incompressibility conditions. The constant symmetric traceless tensors in front of the components' product would become the minimization parameters.}

The surfaces nested in the other ones drop from the equations. Consider the inner surface $S_{in}$  nested inside the outer surface $S_{out}$.
Constant velocity fields inside $S_{in}$ and inside $S_{out}$ have to match at every point of $S_{in}$, after being projected on its normal vector $\vec \Sigma_{in}$. This matching is possible only if these constant velocities are equal, in which case there will be no discontinuity, and thus the inner surface drops from equations. \footnote{We carefully verified this fact for $N$ concentric spheres with spherical ansatz $\Gamma_n = \gamma_n \vec Q\cdot\vec r$. The minimum of free energy requires all $\gamma_n = 0$ except for the largest sphere.}

Therefore, we always assume that the surfaces are not nested.

Outside all surfaces, the flow is purely potential.  
The potential $\Phi$ is a harmonic function with Neumann boundary conditions on each sphere, following from the continuity of normal velocity
\begin{eqnarray}\label{Neumann}
    && \vec v = \vec \nabla \Phi;\\
    &&\vec \nabla \cdot \vec v = \vec \nabla^2 \Phi =0;\\
    && \vec\Sigma_n \cdot\left(\vec \nabla\Phi -\vec C_n\right)_{S_n} =0 ;\\
    && \Gamma_n(\vec r) = \Phi(\vec r) - \vec r \cdot \vec C_n;\, \vec r \in S_n;
\end{eqnarray}

This boundary condition also applies to the open surface bounded by some contour. In that case, the condition involves a constant vector velocity $\vec C_n$, the same one on both sides of the surface. The inner region is absent in this case, or, better to say, it shrinks to a zero thickness layer between the two sides of the surface.

For a potential steady flow in each domain $D_i$ separated by surfaces, the Euler equation is solved for constant total pressure (Bernoulli equation)
\begin{equation}
    P_i = p + \oh \vec v^2 = \mbox{const} ;\;\forall \vec r \in D_i
\end{equation}
To get some restrictions on the constants $P_i$ in different domains $D_i$ of potential flow (inside each surface and outside all of them), let us apply the Stokes theorem to the Euler equation for the  time derivative of energy, without assumptions of global potential flow
\begin{eqnarray}
    &&0 = -\int \val \d_t \val = \int d^3 r  \val \left( \dal p + \vbe\dbe \val\right) = \nonumber\\
    &&\int d^3 r  \val \dal\left(  p + \oh \vec v^2\right) = \sum_n \int_{S_n} d S v_n \Delta_n P =\nonumber \\
    &&\sum_n \Delta_n P\int_{S_n} d S v_n = \sum_n \Delta_n P \int_{D:\d D = S_n} \dal \val =0;
\end{eqnarray}
Here $\Delta_n P$ is the discontinuity of the total pressure at the boundary $S_n$ between two domains of potential flow.
We used the continuity of the normal velocity $v_n = \vec \Sigma \cdot \vec v ; \vec r \in S$. 
We observe that these discontinuities $\Delta_n P$ drop from the energy conservation: it is identically satisfied in virtue of incompressibility. Thus, these constant values of the total pressure are not influencing the vortex sheet dynamics and can all be set to zero.

The net momentum equals to
\begin{equation}\label{MomGamma}
    \vec P = \ot \sum_n\int_{S_n} \Gamma_n d \vec \sigma
\end{equation}
where $d \vec \sigma = d\xi_1 d \xi_2\d_1 \vec X \times \d_2 \vec X$ is vector area element.

Let us assume that we solved the Neumann Laplace problem.
Then we can express  $\Phi(\vec r)$ as linear combination of all the vectors $\vec C_m$ which define  the boundary conditions for the external potential.
\begin{eqnarray}\label{Neumann1}
    &&\Phi(\vec r) = \sum_m \vec C_m \cdot \vec \Psi_m(\vec r);\\
    &&\Gamma_n(\vec r) = \sum_m \vec C_m \cdot \vec \Psi_{m n}(\vec r);\\
    && \vec \Psi_{m n}(\vec r) = \vec \Psi_m(\vec r) - \vec r \delta_{m n}; \; \vec r \in S_n;
\end{eqnarray}
This harmonic vector field  $\vec \Psi_m(\vec r)$ satisfies the Neumann boundary conditions
\begin{eqnarray}\label{Neumann2}
    \left(\vec\Sigma_n \cdot\vec \nabla\Psi^\mu_m\right)_{S_n}  = \Sigma_n^\mu \delta_{n m}
\end{eqnarray}
This field is universal, given the geometry of discontinuity surfaces, 
which gives us the opportunity to minimize the free energy as a quadratic form made of $\vec C_n$.

Substituting this into the momentum equation \eqref{MomGamma} yields the linear equation
\begin{eqnarray}
    &&Q_\mu = \sum_m  M_m^{\mu\nu}C_m^\nu;\\
    && M_m^{\mu\nu} = \ot \sum_n\int_{S_n}d \sigma^\mu \Psi_{m n}^\nu
\end{eqnarray}
The Hamiltonian can also be expressed in these functions
\begin{eqnarray}
&& H = \oh \sum_{m n} C_m^\mu  H_{m n}^{\mu\nu} C_n^\nu;\\
     &&H_{m n}^{\mu\nu}= \sum_{k l} \int_{S_k} \int_{S_l}\frac{ d \Omega_{m k}^{\mu\alpha}(\vec r)d \Omega_{n l}^{\nu\alpha}(\vec r')}{4 \pi | \vec r - \vec r'|};\\
     && d \Omega_{m k}^{\mu\alpha}(\vec r) = d  \Psi_{m k}^\mu(\vec r)\wedge d r^\alpha;
\end{eqnarray}
We have to minimize the free energy
\begin{eqnarray}
    F =\oh\sum_{m n} C_m^\mu  H_{m n}^{\mu\nu} C_n^\nu - \lambda^\mu  \sum_m  M_m^{\mu\nu}C_m^\nu + \lambda^\mu Q^\mu
\end{eqnarray}
after which the vector parameters $\vec C_1 \dots \vec C_n,\vec \lambda$ will be linearly related to  $\vec Q$ and thus $\Gamma_n$ will have a form
\begin{eqnarray}
    \Gamma_n(\vec r)  = \vec \Xi_n(\vec r) \cdot \vec Q;
\end{eqnarray}

We are left with the last problem: to find the momentum $\vec Q$ from the energy balance equation $E[\Gamma,\vec X] = \vec Q  \cdot \vec f$.
We find the following equation:
\begin{eqnarray}
    && Q_\alpha  M^{\alpha\beta} Q_\beta = Q_\alpha f_\alpha;\\
    && M^{\alpha\beta} = \Lambda \sum_n \int_{S_n}  \d_\mu \Xi_n^\alpha \left(\delta_{\mu\nu} - \Sigma_\mu\Sigma_\nu\right)\d_\nu\Xi_n^\beta ;
\end{eqnarray}
where as before, $\vec \Sigma$ is the unit normal vector to the surface.

This is an equation for the momentum $\vec Q$ as a function of the force vector $\vec f$. The relevant nonzero solution is just a $3 \times 3$ matrix inversion:
\begin{eqnarray}
    \vec Q = \hat M^{-1} \cdot \vec f;
\end{eqnarray}
The energy flow after averaging over Gaussian vector $\vec f$ with variance $\sigma$
\begin{eqnarray}
    \VEV{\mathcal{E}} = \VEV{\vec f \cdot \vec Q} = \sigma \tr \hat M^{-1};
\end{eqnarray}

This minimization can be readily done for a sphere with the following results:\footnote{In this case, due to the symmetry, the constant velocity $\vec v(\vec r) =a \vec Q$ inside the sphere is the most general solution, linear in $\vec Q$.}:
\begin{subequations}\label{solsphere}
\begin{align} 
&\Phi(\vec r) = \frac{\vec q \cdot \vec r}{3}\begin{cases}
-2 &\text{for } |\vec r| < R
\\
 R^3/|\vec r|^3&\text{for } |\vec r| > R\text{;}
 \end{cases}
 \\
&\vec v = \vec \nabla \Phi =
    \begin{cases}
-2\vec q/3 &\text{for } |\vec r| < R \text{;}
\newline
\\
-\vec q/6 -\hat r (\hat r \cdot \vec q)/2&\text{for } |\vec r| = R \text{;}
\newline
\\
 \left( \vec q/3 - \hat r (\hat r \cdot \vec q)\right) R^3/|\vec r|^3&\text{for } |\vec r| > R\text{;}
    \end{cases}\\
      & \hat r = \frac{\vec r}{R};\\
  & \Gamma(\vec r) = \Phi^+(\vec r) - \Phi^-(\vec r) = \vec q \cdot \vec r& \text{ for }|\vec r| = R ;\\
  & \vec q  = \frac{2 R^5}{3 \Lambda} \vec f;\\
 & \vec P =\frac{ 16 \pi R^4 \vec f}{27\Lambda};\\
&\sigma = \frac{9 \mathcal E \Lambda}{16 \pi R^8};\\
&p(\vec r) = \textit{const } - \oh \vec v^2
\end{align}
 \end{subequations}
The coefficients $-2,1$ for $\Phi$ inside and outside follow from the Neumann boundary condition for radial derivative of $\Phi$.
The computation of $q, \vec P, \sigma$ is performed in Appendix A.
This solution is manifestly gauge invariant, as we expressed $\Gamma$ as a function of a point $\vec r$ in 3D space, projected on a surface, without specifying the parametrization of the surface. 

It is a good problem for a grad student to perform the minimization of the Hamiltonian \eqref{Ham} for two concentric spheres with ansatz $\Gamma_n = \vec q_n \cdot \vec r$ at each sphere. They should verify that the smaller one drops from the solution (has zero $\vec q_1$).
Everything looks different in 2D, but it is guaranteed to come out the same as we have just found in the 3D Laplace equation because this is a linear minimization problem with a positive non-degenerate quadratic form.

\section{Viscosity Anomaly and Scaling Laws}

Balancing the terms in the energy flow equation in the turbulent limit ($\nu \ra 0, \mathcal{E} = \mbox{const}$) led us in\cite{M20c} to new scaling laws, different from Kolmogorov scaling law.

Let us repeat these arguments now, with our new understanding of the steady vortex sheets. We also reproduce in our solution the Gaussian profile \cite{M20c} of vorticity in the viscous layer around Euler \DS{}.

Actually, this solution of the \NS{} equations was known for quite some time\cite{BM05}, being discovered first by Burgers\cite{BURGERS1948}. In our notation it reads
\begin{subequations}\label{BurgeSht}
\begin{eqnarray}
  &&\vec v = \{ - a x, b S_h(z), a z\};\\
  &&\vec \omega =\{-b S_h'(z),0,0\} ;\\
  &&S_h'(z) =\frac{2}{h \sqrt{2 \pi} } \EXP{- \frac{z^2}{2 h^2}};\\
  && S_h(z) =\erf \left( \frac{z}{h \sqrt 2}\right);\\
  && a = -\frac{\nu}{h^2};\label{aheq}
\end{eqnarray}
\end{subequations}
This solution has two free parameters $h, b$.

In the limit of small $h$  we have the vortex sheet with constant tangent velocity discontinuity
\begin{eqnarray}
    &&S_0(z) = \sign(z);\\
    &&S_0'(z) = 2 \delta(z),
\end{eqnarray}
so that this solution leads to the viscosity anomaly, as we shall shortly see.  

Before we do that, let us clarify the terms. The persistence of dissipation in the \NS{} equation was known for quite some time -- this is the core of the turbulence phenomenon. 

Various great researchers in the past were looking for the dynamical mechanism of this anomalous dissipation. 

The Richardson cascade and Kolmogorov scaling were based on the presumption that this persistent dissipation occurred because of the growing number of small eddies as the energy  cascades from large spatial scales to smaller ones.

This hierarchical cascade tree says nothing about the spatial distribution of small eddies.  As we claim now, in a strong turbulence limit, they collapse in vortex sheets, just as it was foreseen by Burgers.

Let us balance the powers of viscosity in our equations,
assuming that the width $h$ of this layer goes to zero as some power of $\nu$.

We could have three scaling laws
\begin{subequations}
\begin{eqnarray}
    &&h \propto \nu^\alpha;\\
    &&\Phi ,\vec v, \vec \omega, a, b , \Gamma \propto \nu^\beta;\\
    &&\vec f \propto \nu^\gamma;
\end{eqnarray}
\end{subequations}

We shall find the unknown powers from the balance of energy and the equations of motion. All surfaces are finite so that the coordinates $\vec X$ are not supposed to scale with viscosity. In other words, we only compare the powers of time in our stationary equations. The only source of such scale factors of time is the viscosity.

We consider the local tangent plane to the vortex sheet and in the linear vicinity $x,y,z \ra 0$ we could use the Burgers solution \eqref{BurgeSht}. 
Let us note that this is \textbf{not} how we presumed the \NS{} equations to be satisfied inside the viscous layer in our previous work\cite{M20c}. We did not have the term $v_x = -a x$ there and we got a correct Gaussian profile just by matching the singular terms. 

Fortunately, the old work by Burgers\cite{BURGERS1948} gives us an exact solution of the \NS{} equation in the linear vicinity of the tangent plane without any further assumptions. We can see now how all regular terms in the \NS{} equation cancel as well as the singular terms.

Outside that boundary layer, we have a purely potential steady flow, solved by the Bernoulli equation inside each surface or outside all of them.

These potential flows are matched at each surface, with the normal velocity continuous and local tangent discontinuity matched by the Burgers solution in the local tangent plane inside each viscosity layer of vanishing width $h \sim \nu^{\thfi}$.
In the turbulent limit, $h \ra 0$ error function becomes the sign function, and we recover our tangent discontinuity. 

This solution applies to the curved surface with $x,y$ being the coordinates in a local tangent plane assuming that the surface curvature is less than $h^{-2}$. This simple fact was not fully realized by Burgers and his followers, who did not understand the universal nature of his anomalous dissipation.

We also see that the velocity discontinuity in this solution in the limit $h \ra 0$
\begin{eqnarray}
    \Delta \vec v = \{0, 2b, 0\}
\end{eqnarray}
This corresponds in our notation to the linear $\Gamma$
\begin{equation}
    \Gamma \ra 2b y
\end{equation}

Naturally, we can choose the direction of the $y$ axis as we wish, so this corresponds to directing it along $\vec q = \vec \nabla \Gamma$.
In general, we would have to rotate the Burgers solution in the $x y$ plane by some constant angle $\theta$ to align the $y$ axis with local $\vec \nabla \Gamma$. We can also translate the velocity in the $ x,y$ plane and add the corresponding linear term to the pressure.

Here is this translated and rotated solution, generated by the \Mathematica{} code\cite{MB},which also checks that it satisfies the \NS{} equation:
\begin{subequations}\label{rotBurg}
\begin{align}
& a = - \frac{\nu}{h^2};\\
   &\vec v = \begin{Bmatrix}
   &-a \cos (\theta ) (x \cos (\theta )+y \sin (\theta ))-b \sin (\theta ) S_h(z)+c\\
   & -a \sin (\theta ) (x \cos (\theta )+y \sin (\theta ))+b \cos (\theta ) S_h(z)+d\\
   &a z
   \end{Bmatrix};\\
   & p = -\frac{1}{2} a^2 \left((x \cos (\theta )+y \sin (\theta ))^2+z^2\right)\nonumber\\
   &+a x \cos (\theta ) (c \cos (\theta )+d \sin (\theta ))+a y \sin (\theta ) (c \cos (\theta )+d \sin (\theta ));\\
   & \vec q = \Delta \vec v = \{-2 b \sin (\theta ),2 b \cos (\theta ),0\};\\
   & \Gamma = \vec q \cdot \vec r;
\end{align}
\end{subequations}

In the case of our spherical solution in the Galilean frame $\vec v \Ra \vec v + \tt{} \vec q$ where the normal velocity component vanishes at the surface, we have the vector $\vec q$ given, and the velocity at the middle 
\begin{eqnarray}
    &&\vec v(|\vec r| = R) = \oh \left(\vec q -\hat r \vec q \cdot \hat r\right);\\
    && \hat r = \frac{\vec r}{R};
\end{eqnarray}

The positive parameter $h$ remains arbitrary in the Burgers solution.  This is a universal free parameter of our theory, related to the dissipation factor $\Lambda$.

Now we can match the transverse components of velocity  with the generalized Burgers solution and find the parameters $\theta, b, c, d$ 
in a local frame where $\vec r = \{0,0,R\}$
\begin{eqnarray}
    &&\{c,d\} = \oh\{q_x, q_y\}\\
    && \{- b \sin (\theta ), b \cos (\theta )\} = \oh\{q_x, q_y\};\\
\end{eqnarray}
We solve these equations for $b, \theta$ to find finally
\begin{eqnarray}
    && b = \oh |\vec q|;\\
    && \theta = -\arcsin\frac{q_x}{|\vec q|}.
\end{eqnarray}

This solution assumes we are on the north pole of the sphere. Now we need to rotate the Burgers solution so that the north pole transforms into a given point on a sphere.

We have to find rotation matrix $\Omega(\{x,y,z\})$ such that unit vector $\vec r/|\vec r| = \{x,y,z\}$  transforms into the the vector $\{0,0,1\}$ on a sphere. We also have to shift the coordinate origin from the center of the sphere to its surface by subtracting $\vec r$ from the origin.

This rotation matrix is found in our \Mathematica{} code\cite{MB}
\begin{align}
    &\Omega(\{x,y,z\}) = \left(
                \begin{array}{ccc}
                 \frac{x^2}{1-z}-1 & \frac{x y}{1-z} & -x \\
                 \frac{x y}{z-1} & \frac{y^2}{z-1}+1 & y \\
                 x & y & z \\
                \end{array}
                \right);\\
          &\Omega(\hat r)\cdot \hat r = \{0,0,1\};\\
          &\vec r_\Omega = \Omega(\vec r/|\vec r|) \cdot \vec r -\{0,0,R\};
\end{align}

We use this rotation with constant $\{x,y,z\}= \hat r_0$ and take the Burgers solution $\vec v_B(\delta \vec r)$ in its linear vicinity
\begin{eqnarray}
   &&\Omega_0 = \Omega(\hat r_0);\\
   && \delta \vec r = \Omega_0 \vec r - \{0,0,R\};\\
   &&\vec v( \vec r) = \Omega_0 \vec v_B(\delta \vec r);\\
   && p( \vec r) = p_B(\delta \vec r).
\end{eqnarray}

As the \NS{} equation is invariant with respect to translation and rotation with constant parameters, this transformed Burgers solution will also satisfy the \NS{} equation.

This completes the matching of the Burgers vortex sheet with our Euler vortex sheet solution.

Matching powers of $\nu$ in the \eqref{aheq}, we find the first relation
\begin{equation}
    \beta =  1- 2 \alpha
\end{equation}
We have to assume that all terms of the velocity scale as the same power of viscosity, to fulfil the rotational invariance
\begin{eqnarray}
    b \sim a \sim \nu^{\beta},
\end{eqnarray}

The second relation follows from the forced energy pumping
\begin{eqnarray}
    \mathcal{E} = \vec f \cdot \vec P \sim \nu^{\gamma + \beta};
\end{eqnarray}

Demanding finite $\mathcal{E}$ we find the second relation
\begin{equation}
     \gamma = -\beta
\end{equation}

Finally, the energy dissipation
\begin{eqnarray}
    \mathcal{E} = \nu \int d^3 r \vec \omega^2 \ra \nu \int_{-\infty}^{\infty}  d z \int_S d^2 r  (\d_z \vec v)^2 
\end{eqnarray}
We leave only normal derivatives of velocity here, as the other components would not lead to a large factor to compensate $\nu$ in front.
The normal derivatives of the velocity produce a Gaussian function of the local normal coordinate $z$ which after integration yields
\begin{eqnarray}
    &&\nu \int_{-\infty}^{\infty}  d z \int_S d S (\d_z \vec v_t)^2\ra \nonumber\\
    &&\frac{\nu}{2 \pi h^2 }\int_{-\infty}^{\infty} d z  \EXP{-\frac{z^2}{h^2}} \int_S d S (\vec \nabla\Gamma)^2\nonumber\\
    && = \frac{\nu }{2 h \sqrt{\pi}}\int_S d S (\vec \nabla\Gamma)^2 \propto \nu^{1-\alpha + 2 \beta}
\end{eqnarray}
Demanding a finite $\mathcal{E}$ once again, we find the third  relation
\begin{equation}
    1-\alpha + 2 \beta =0
\end{equation}

Solving these three equations we find (just as in \cite{M20c}, but with all terms in the \NS{} equation matched now):
\begin{subequations}
\begin{eqnarray}
    && \alpha = \frac{3}{5};\\
    && \beta = -\frac{1}{5};\\
    && \gamma = \frac{1}{5}.
\end{eqnarray}
\end{subequations} 

This observation leads us to believe that we have found the steady solution of the full \NS{} equation in the turbulent limit. 
In the boundary layer around each vortex surface, we have the rotated Burgers solution \eqref{rotBurg} in the local tangent plane, matching the local velocity discontinuity $\Delta \vec v = \vec \nabla \Gamma$. Outside, we solved for the steady potential flow using Bernoulli equation and matching normal derivatives at each surface.

Note that the powers match in our exact spherical solution \eqref{solsphere}, as they should:
\begin{eqnarray}
    && \Lambda \sim \frac{\nu}{h} \sim \nu^{1-\alpha};\\
    && \vec v \sim \vec Q \sim \frac{\vec f}{\Lambda} \sim \nu^{\beta};\\
    &&\vec f \sim  \sqrt \sigma \sim \nu^{\gamma};
\end{eqnarray}

Positive $\alpha$ justifies our assumption of the viscous layer $h$ shrinking to zero in the turbulent limit and the error function becoming the sign function.

Positive $\gamma$ means that the external force goes to zero in the turbulent limit, but a large value of velocity field compensates that in energy pumping. In enstrophy, the large factor $1/\nu$ comes from the large square of vorticity times the small width of the vorticity layer. The resulting large factor compensates the factor of $\nu$ in front of the enstrophy leading to finite energy dissipation.

Therefore, just like in the critical phenomena, the infinitesimal external field is enhanced by a large susceptibility. The susceptibility is large due to the singular vorticity coming from large gradients of velocity in the viscous layer surrounding the Euler discontinuity surface.

This enhancement makes these vorticity sheets the dominant configuration in the turbulent limit, responsible for the energy dissipation.

The reader must have a natural question: what about the K41 scaling, which dominated the turbulence theory for half a century? It became even more complicated in the last 30 years, with multi-fractal scaling laws -- nothing like simple rational indexes.

The real answer is that our scaling laws say nothing about the energy spectrum or spatial dependence of velocity/vorticity correlation functions.

Moreover, these new scaling laws correspond to extreme turbulence, which is just beginning to reveal itself in DNS. We are planning large-scale DNS in collaboration with K.Iyer, to verify the predictions of vortex sheet theory.

We are also working with Nigel Goldenfeld and Dmytro Bandak on the explanation of an old experiment where the planar turbulent vortex sheet was created and the PDF of its dissipation was studied. 

For some mysterious reason, it matches the distribution of magnetization of the 2D spin wave model with temperature going to zero as Reynolds to the power of $-\ot$.

Later in this paper, we mention why it is natural to expect such an analogy in our Gibbs statistics of vortex sheets.

\section{Topological Invariants}

Let us now compare this system's topology with the one discussed in our recent review paper \cite{M20c}.

The simplest case is the one where all surfaces $S_n$ are closed.

In that case, we can introduce  Clebsch field $\phi_1(\vec r),\phi_2(\vec r)$ as usual
\begin{equation}
    \vec \omega = \vec \nabla \phi_1 \times \vec \nabla \phi_2
\end{equation}

Then, the second field $\phi_2$ taking constant values inside each closed surface and zero outside would lead to our ansatz with $\Gamma = \phi_1 \Delta \phi_2$. The values of the \CL{} field $\phi_1$ outside the surface drops from the equation. One can take any smooth interpolating field $\phi_1$ between these surfaces, and the vorticity field will stay zero outside and inside the surfaces.

Furthermore, the velocity circulation around any contractible loop at each surface vanishes because there is no normal vorticity on these surfaces.

The \CL{} topology plays no role in this case, unlike the case with open surfaces with edges $C_n = \d S_n$ we considered in \cite{M20c}.
There, the normal component of vorticity at the surface was present (and finite).

Still, our collection of closed vortex sheets in the general case has some nontrivial helicity \eqref{Hel}. This helicity is pseudoscalar, but it preserves the time-reversal symmetry. 

The parity transformation $P$ changes the sign of velocity, keeping vorticity invariant, whereas the time-reversal $T$ changes the signs of both velocity and vorticity.

\begin{figure}
    \centering
    \includegraphics[width=\textwidth]{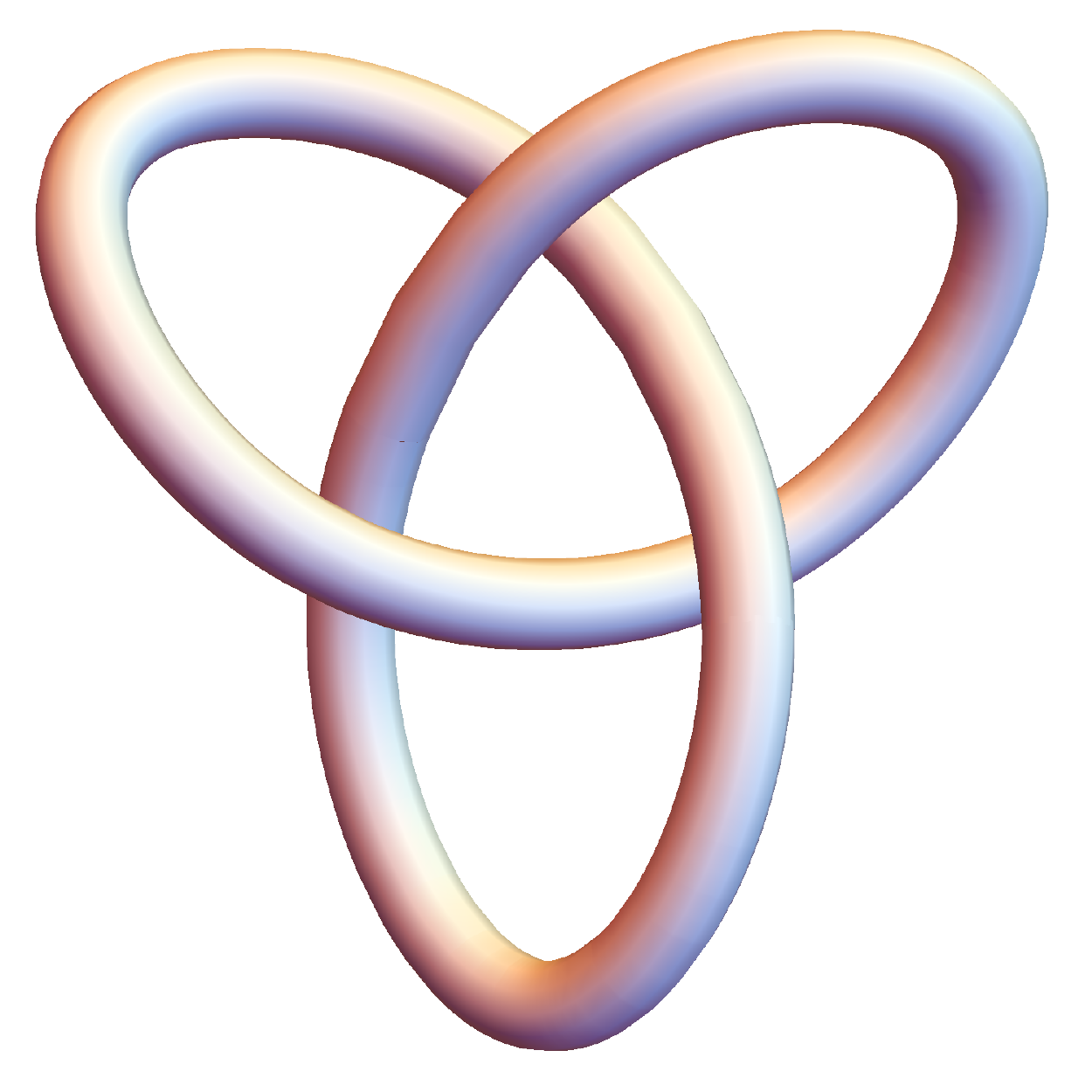}
    \caption{The torus knotted to produce nontrivial helicity.}
    \label{fig::Trefoil}
\end{figure}

The helicity integral is $T$-even and $PT$-odd. It measures the knotting of vortex lines between these surfaces. 
(see Fig.\ref{fig::Trefoil}). 
As it was noted in \cite{AM89} the surfaces avoid each other and themselves, so these are not just some random surfaces. This property was studied in their time evolution and recently in their statistics \cite{M20c}.

The circulation around each contractible loop on the surface will still be zero, but the loop winding around a handle would produce a topologically invariant circulation  $n \Delta \Gamma$ for any loop winding $n$ times.

This $\Delta \Gamma$ is the period of $\Gamma(\xi)$ when the point $\xi$ goes around this handle\cite{M88}. This period depends upon the surface's size and shape, but it does not change when the path varies along the surface as long as it winds around the handle. 

There is a flux through the handle related to tangential vorticity inside the skin of the surface. This flux through any surface intersecting the handle is topologically invariant, and it equals to $\Delta \Gamma$.

If we consider the circulation around some fixed loop in space, it will reduce to an algebraic sum of the circulations around all closed surfaces' handles encircled by this loop. It will be topologically invariant when the loop moves in space without crossing any of the surfaces. 

\begin{figure}
    \centering
    \includegraphics[width=\textwidth]{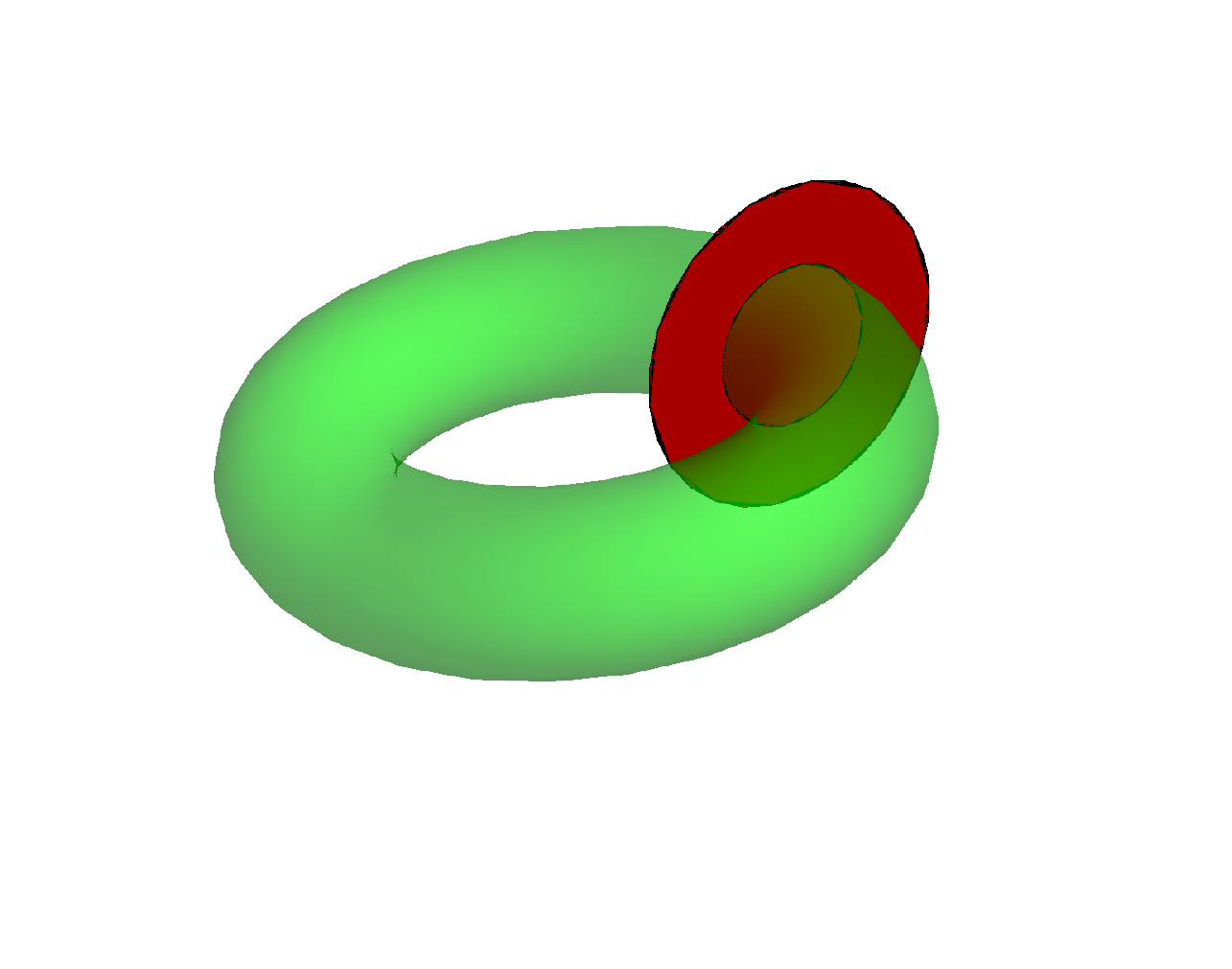}
    \caption{The green vortex tube $T$ cut  vertically by a red disk $D_C$. The vorticity flux through the disk reduces to integral of velocity discontinuity over $\alpha$ cycle.}
    \label{fig::CutTorus}
\end{figure}

In particular, the closed vortex tube (topological torus) encircled by a fixed loop $C$ in space would produce the circulation $\oint_C \vec v \cdot d \vec r = \Delta \Gamma$ which is equal to the period of $\Gamma$ around the $\alpha$ cycle, corresponding to shrinking of $C$.
(see Fig.\ref{fig::CutTorus}). 

\begin{figure}
    \centering
    \includegraphics[width=\textwidth]{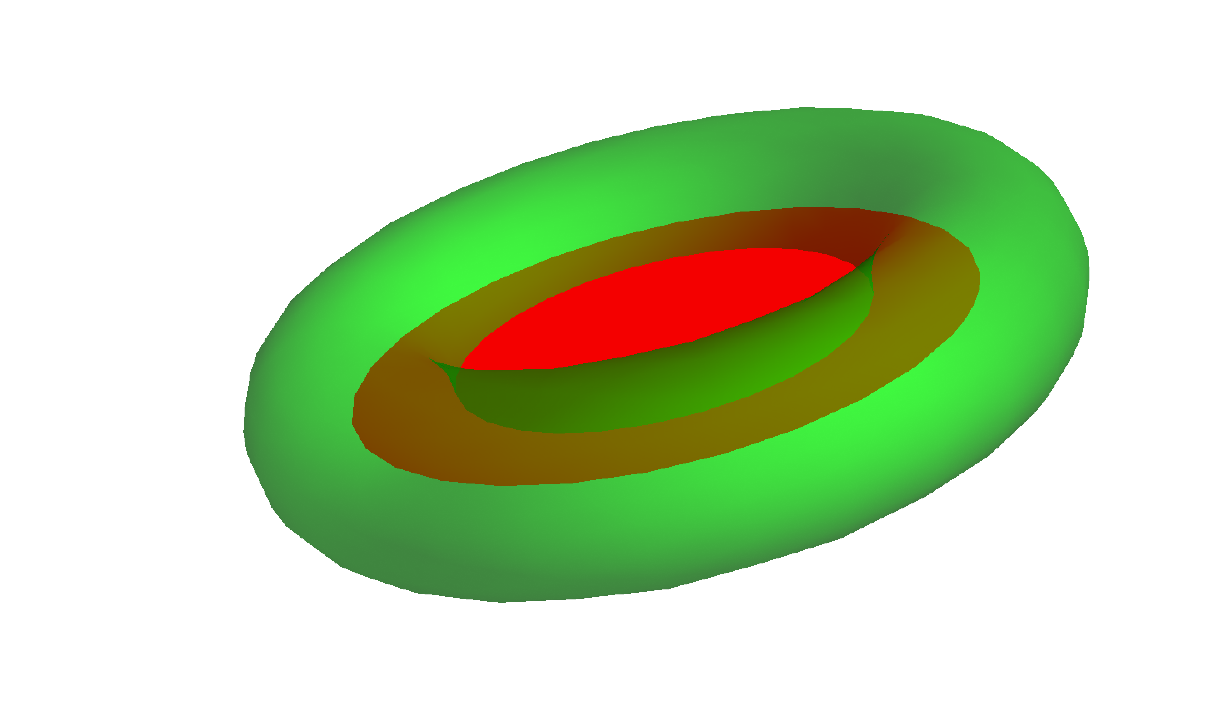}
    \caption{The green vortex tube $T$ cut horizontally by a red disk $D_C$.The vorticity flux through the disk reduces to integral of velocity discontinuity over $\beta$ cycle.}
    \label{fig::HCutTorus}
\end{figure}

Another cross-section of the same vortex tube leads to circulation around another cycle of the torus.
(see Fig.\ref{fig::HCutTorus}). 

To prove the relation between circulation over the disk edge $C$ and $\Delta \Gamma$ over the corresponding cycle of the handle, let us consider the flux of vorticity through the disk $D_C$ bounded by $C$ intersecting this torus $T$.

By the Stokes' theorem, this flux is equal to the circulation on the external side of $T$ minus circulation on the internal side. On the other hand, it is equal to the circulation around the edge $C$ of this disk
\begin{eqnarray}
    \oint_C \vec v \cdot \vec d r = \oint_{\gamma\in T}  \Delta \vec v d \vec r = \oint d\Gamma = \Delta \Gamma
\end{eqnarray}

\section{Turbulent Statistics}

According to the approach to turbulent statistics that we presented in\cite{M20c}, the Hopf equation's fixed point corresponds to a steady flow with random initial data and Gaussian random force $\vec f$.

The Hopf equation\cite{Hopf19} represents a functional equation for generating functional $Z[\Phi]$ for the distribution of various variables $\Phi$ like velocity, vorticity, or Clebsch variables. The Hopf equation follows directly from the dynamics, assuming some distribution of initial and boundary data. It has a form
\begin{eqnarray}
    \d_t \mathcal Z[\Phi] = \hat H\left[\ff{\Phi}\right] \mathcal Z[\Phi]
\end{eqnarray}

The stationary distribution $\mathcal Z^*[\Phi]$ would correspond to a fixed point of this equation.
\begin{eqnarray}
    \hat H\left[\ff{\Phi}\right] \mathcal Z^*[\Phi] =0
\end{eqnarray}

What is the random distribution of parameters in our stationary solution? After expressing the function $\Gamma_n$ at each surface in terms of the shape $\vec X_n(\xi)$  of these surfaces by minimizing the Hamiltonian, we are left with this $ \vec X_n(\xi)$ as the initial data to randomize.

On a second thought, we rather keep the quadratic Hamiltonian \eqref{Ham} with large coefficient $\beta$ (effective inverse temperature) and study the Gibbs distribution with both $\Gamma(\xi), \vec X(\xi)$ fields present. 

In that case, we have the Hamiltonian residual value at the steady manifold $\Gamma = \Gamma^*[\vec X]$ as an effective Hamiltonian for the surface degrees of freedom.

This approach seems like a natural choice for the effective Hamiltonian for these surface degrees of freedom, which was left arbitrary at this point. All we know so far is that there is a degenerate stationary solution of Euler dynamics involving arbitrary surfaces $\vec X_n(\xi)$ and corresponding $\Gamma_n $ minimizing our Hamiltonian.

We expect this solution to be unstable, but this is the whole point of the Gibbs distribution. For arbitrary initial values of these surfaces, assuming steady values of $\Gamma^*$, we would have the surfaces evolve while preserving their topology and avoid each other and themselves. Eventually, they cover some manifold corresponding to a fixed point of the Hopf equation.

The idea of an effective Gibbs distribution for the velocity field in turbulence was expressed long ago by Onsager\cite{Onsager1949}. More recently, we also speculated about Gibbs distribution of closed vortex cells in \cite{TSVS}, using the symplectic measure of the \CL{} variables and suggesting some effective Hamiltonian from the symmetry principles. 

The new level we have reached now is the specific form of the Hamiltonian and some microscopic dynamics leading to that Hamiltonian.

\textbf{We are now suggesting that the turbulence is the Gibbs distribution of vortex sheets, with some external field terms in the effective Hamiltonian, supporting the energy flow. Great advantage over previous approaches is that this vortex sheet statistics is a 2D theory, exactly solvable in the turbulent limit}.

We know that Gibbs distribution $\EXP{- \beta H[\Gamma,\vec X]}$ represents a fixed point of the Hopf equation, because formally the Hamiltonian is conserved as well as the phase space measure $D\Gamma D V[\vec X]$. The ergodicity hypothesis, which we imply here, says that this is a stable fixed point. We modify the Hamiltonian to compensate for the anomalous dissipation inside these vortex sheets.

The real physical fluids always have finite temperature (except the quantum superfluids). Therefore, they are always distributed by a Gibbs distribution, but lately the Gibbs distribution was discarded as the origin of the turbulent statistics.

The thermal fluctuations of velocity in the 3D approach merely add to the turbulent ones at the viscous scales and below. See the recent work\cite{eyink20}, where the idea of spontaneous stochasticity was promoted and the role of thermal fluctuations was discussed.

The thermodynamics of the laminar fluid (large viscosity limit) involves different mechanisms. In that case, there are no vortex sheets, and the description in terms of Wylde functional integral with weakly fluctuating velocity field would be an appropriate approach. It is no longer a statistics, but rather a time-dependent process, weak turbulence, described by kinetic equations\cite{YZ93}.

It is just the strong turbulence phase with vortex sheets where the Gibbs distribution with energy flow/dissipation constraints is applicable. For the vortex sheet statistics, strong turbulence represents a weak coupling (low-temperature) phase.

The measure in the phase space of $\Gamma, \vec X$ corresponds to canonical pair $\Gamma, V$ in our action \eqref{action}.
\begin{eqnarray}
   && d\mu[\Gamma, \vec X] = d\Gamma d V[\vec X] ;\\
   &&d V [\vec X] = d \vec \Sigma \cdot d \vec X
\end{eqnarray}
Here $V$ is the volume inside the surface. Its variation $d V$ is proportional to the  displacement $d \vec X$ projected by area element vector $ d\vec \Sigma = d^2 \xi \d_1\vec X  \times \d_2 \vec X$.

Note that the tangent displacements of $\vec X$ which just re-parametrize the surface without changing its volume, are eliminated here so that only normal displacement $d X_\perp$ is left.

As usual in gauge theory, some gauge conditions could be added, eliminating this re-parametrization and factoring out the volume of this gauge orbit.

These issues were analyzed in great detail in string theory as a theory of random surfaces, starting with the pioneering work\cite{Pol81}. Here we have an additional term in the effective Hamiltonian, proportional to the volume $V$ inside the surface.

Another new aspect is the requirement that this surface is self-avoiding \cite{M88,AM89,M20c}.

Assuming the Gibbs distribution with energy flow constraints, we can advance this distribution further.

Let us introduce two Lagrange multipliers $\lambda, \rho $ for fixing energy pumping and energy dissipation in the Gibbs microcanonical partition function
\begin{eqnarray}
    && \int D\Gamma D V[\vec X]\EXP{ - \beta H[\Gamma, \vec X ]}\nonumber\\
    &&\int d \lambda d\rho\EXP{\i \rho \left(\mathcal{E} - E[\Gamma,\vec X]\right)}\nonumber\\
    && \int d^3 f \EXP{-\frac{\vec f^2}{2 \sigma}+\i \lambda \left(\mathcal E - \vec f \cdot \vec P[\Gamma,\vec X]\right)};
\end{eqnarray}
with $\vec P[\Gamma,\vec X]$ given by \eqref{PGam} and $E[\Gamma,\vec X]$ by \eqref{EGam}. 

These two constraints compensate for the lack of energy conservation in our dynamics. The time derivative of our distribution involves
\begin{eqnarray}
    \d_t H[\Gamma, \vec X] = \vec f \cdot \vec P[\Gamma, \vec X] - E[\Gamma,\vec X].
\end{eqnarray}
which vanishes due to our constraints.

\section{Dissipation of Dissipation}

For the conserved distribution in the sense of the Liouville theorem for the Gibbs statistics, the constraints themselves must be conserved in the Euler dynamics. This is straightforward in case of the pumping term, as the momentum $\vec P[\Gamma, \vec X]$ is, indeed, conserved, and the force $\vec f$ is constant.

In case of dissipation, we need to actually check  its conservation.
It is easier to start from its original definition as an enstrophy and differentiate it using \NS{} equations
\begin{eqnarray}
    &&\d_t \nu \int d^3 r\vec \omega^2 = 2\nu \int d^3 r \oal \d_t \oal =\nonumber\\
    && 2\nu \int d^3 r \vbe \dbe \left(\oh \oal^2\right) - \oal  \obe  \dbe \val 
\end{eqnarray}
Integrating the first term by parts and ignoring the boundary terms at infinity (as $\omega =0$ there), we are left with the second term
\begin{eqnarray}
    && -2 \nu\int d^3 r \oal \obe \dbe \val
\end{eqnarray}

This integral can be computed in the turbulent limit in the same way as we did with the dissipation itself. Using the asymptotic solution in vicinity of the surface, we find (in the local tangent plane frame)
\begin{eqnarray}
   && \d_t E[\Gamma, \vec X] = -2 \Lambda \int d S  \tilde \d^i \Gamma \tilde \d^j \Gamma \d_i v_j;\\
   && \tilde \d^i = e^{i k} \d_k
\end{eqnarray}
Here $v_j$ are the tangent components of the velocity field, understood, as usual in vortex dynamics, as the average of the velocity inside  and outside the surface.
Note that the most singular terms with $(\delta(z))^3$ cancelled here, as the vorticity is tangential and tangent derivatives of velocity do not contain $\delta(z)$.

Furthermore, as we have seen in the previous sections, the solution for velocity inside the surface is a constant vector, proportional to the conserved total momentum. Therefore, the gradient of the average velocity reduces to one half of the gradient of the velocity gap.
\begin{eqnarray}
    \d_i v_j = \oh \d_i \d_j \Gamma
\end{eqnarray}

After this, we get 
\begin{eqnarray}\label{dtE}
   \d_t E[\Gamma, \vec X] = -\Lambda \int d S  \tilde \d^i \Gamma \tilde \d^j \Gamma \d_i \d_j \Gamma
\end{eqnarray}

This expression does not vanish in general, which forces us to conclude that the dissipation is, in turn, also  dissipating in our anomalous vortex sheet dynamics.

We have the situation already encountered in the dynamics of constrained systems. The first set of constraints does not commute with the Hamiltonian in the sense of Poisson brackets, so the secondary constraints are generated to compensate for that.  This process can continue until the constraints are conserved at some level, otherwise we have to deal with an infinite number of constraints at all levels.

We leave this interesting complication for the future study, while we work with the truncated theory at the first level of constraints, in the hope that it will prove to be consistent by another reason.

One such reason is that this dissipation of dissipation has two more derivatives than the dissipation itself.
Therefore, in the low wavelength region (inertial range of turbulence)  the next level constraints can be neglected.

The leading terms with just second derivatives correspond to a two-dimensional conformal theory, with the higher derivative terms being so-called irrelevant conformal operators of dimension greater than the dimension $D=2$ of our space.

The experts of conformal field theory will understand the meaning and significance of this observation.

\section{Low-temperature Expansion} 

Let us get back to the Gibbs statistics of vortex sheets with extra constraints for pumping, dissipation and volume inside surfaces.

The next obvious step is to replace the finite-dimensional integration $\lambda, \rho, \vec f$ in the thermodynamic limit of multiple vortex surfaces by saddle points on the imaginary axis. Look at Appendix B in \cite{M20c} for justification of this replacement. 

We need a third Lagrange multiplier -- for the volume \eqref{Vol} inside the closed surfaces, conserved in Euler dynamics.
This gives a term $c V[\vec X]$ in the exponential, where
\begin{eqnarray}
    V[\vec X] = \frac{1}{3}\sum_n \int_{S_n} d^2 \xi \vec X \cdot e_{i j}\d_i \vec X \times \d_j \vec X;
\end{eqnarray}

We arrive at a slightly modified Gibbs canonical ensemble
\begin{align}
    & \mathcal Z(\beta,a,b,c, \vec f)=\nonumber\\
    &\int D\Gamma D V[\vec X]\EXP{- \frac{\vec f^2}{2 \sigma} -\beta H^{eff}[\Gamma, \vec X]};\\
    & H^{eff}[\Gamma, \vec X ] = \nonumber\\
    &H[\Gamma, \vec X ] +a \vec P[\Gamma,\vec X]\cdot \vec f+ b E[\Gamma,\vec X]  + c V[\vec X];\\
    & -\pbyp{\log \mathcal Z(\beta,a,b,c)}{\vec f} =0;\\
    & -\pbyp{\log \mathcal Z(\beta,a,b,c)}{a} = \beta \VEV{\mathcal{E}};\\
    & -\pbyp{\log \mathcal Z(\beta,a,b,c)}{b} = \beta \VEV{\mathcal{E}};\\
    & -\pbyp{\log\mathcal  Z(\beta,a,b,c)}{c} = \beta \VEV{V};
\end{align}
Note that this distribution is Gaussian in terms of $\Gamma$ with quadratic part  being essentially a 2D free massless field kinetic energy $E[\Gamma,\vec X]$. The new soft field appearing from the viscosity anomaly and dominating the turbulent statistics represents the main result of this work.

The nonlocal interaction is provided by  \eqref{Ham}, \eqref{PGam}.

One can remove this nonlocal interaction by introducing a vector Gaussian field
\begin{eqnarray}
   &&\EXP{- \beta H[\Gamma, \vec X ]} \propto \int D \vec \Psi \EXP{-\beta H_1[\Gamma,\vec \Psi, \vec X]}\\
   && H_1[\Gamma,\vec \Psi, \vec X] =\oh \int d^3 r (\vec \nabla \Psi^\alpha)^2 + \i \sum_n \int_{S_n} d \vec \Omega_n^\alpha  \Psi^\alpha;
\end{eqnarray}

Now we have  a local theory of a free 2D field $\Gamma$ interacting with free vector 3D field $\vec \Psi$ (stream function in terms of Fluid Dynamics). The only nonlinear part is the interaction with the surface field $\vec X$.

This functional integral is well defined as the effective Hamiltonian $H^{eff}[\Gamma, \vec X]$  grows at large surfaces as well as large $\Gamma$. The Hamiltonian $H[\Gamma, \vec X ]$ as well as momentum $\vec P[\Gamma,\vec X]$ are manifestly steady in our dynamics with action \eqref{action}, and so is the Liouville measure $D\Gamma D V[\vec X]$. 

The energy dissipation $E[\Gamma,\vec X]$ seems to present a problem as it is not steady in our dynamics. We have computed its time derivative in the previous section, and as a functional of $\Gamma, \vec X$ it does not vanish. We argued that the corresponding extra terms needed to compensate for the  time derivative $\d_t E$ in \eqref{dtE} are higher order in spatial derivatives and as such they can be neglected in the scaling region. Thus, our effective Hamiltonian is steady up to (irrelevant) higher derivative terms.

The Gaussian functional integral $\int D\Gamma$ involved in the partition function reduces to 
\begin{eqnarray}
    &&\mathcal Z(\beta,a,b,c) \propto \nonumber\\\label{XW}
    &&\int D V[\vec X] \frac{\EXP{- \beta H^{eff}[\Gamma^*[\vec X],\vec X]}}{\sqrt{\det{ \hat Q}}};\\
    && \hat Q_{n m}(\xi,\eta) = \frac{\delta^2 H^{eff}}{\delta \Gamma_n(\xi) \delta \Gamma_m(\eta)};
\end{eqnarray}

Now, here is an important detail, which we did not mention so far. As we estimated, the turbulent limit of small viscosity at a fixed energy flow $\mathcal E$ corresponds to large values of $ \Gamma^* \propto \Lambda^{-\oh} \propto\nu^{-\frac{1}{5}} \ra \infty$.

On the other hand, the deviations $\delta \Gamma = \Gamma - \Gamma^*$ are controlled by our effective temperature $\delta \Gamma \sim \beta^{-\oh} \ll \Gamma^* $. We have a WKB situation, where the fluctuations are small compared to the background variable, minimizing the Hamiltonian.

In the zeroth approximation we can neglect these fluctuations and we arrive at the effective $\beta_{eff} \sim \beta  \nu^{-\frac{2}{5}} \ra \infty$. This happens because the leading term in the Hamiltonian $H[\Gamma^*,\vec X] \sim \nu^{-\frac{2}{5}}$.
Therefore, our effective temperature in the Gibbs distribution goes to zero as $T_{eff} = 1/\beta_{eff} \sim \nu^{\frac{2}{5}} \ra 0$.

The temperature will become small only in the limit of ultrahigh Reynolds number $\mbox{Re}\sim \Gamma/\nu \sim \nu^{-\frac{6}{5}}$. We have to wait until $ \beta_{eff} \sim \mathcal \mbox{Re}^{\frac{1}{3}}$ becomes large.

This relation between the effective temperature of the turbulent vortex sheet and Reynolds number perfectly matches the empirical fit of the data, suggesting an analogy between 3D turbulence and spin waves in 2D \cite{Aji_2001}. We interpret these observations as confirmation of our low-temperature Gibbs statistics, with $\delta\Gamma$ playing the role of spin waves in the 2D XY model.

At low temperatures, the saddle point in $\vec X$ (ground manifold) serves as the zeroth approximation. The optimal configuration for the closed surface would be a sphere, and for the open ones bounded by fixed loops, it would be a minimal surface\cite{M20c}.

In the low-temperature expansion, the fluctuations around these minimal surfaces $\delta \vec X \sim \sqrt{T_{eff}} \sim \nu^{\frac{1}{5}}$ are small, so the conditions of self-avoiding are not used. With Gaussian approximation, we shall have a standard perturbation expansion around the ground manifold.

This expansion is opposite to the one in the Wylde functional integral. In our dual theory of random surfaces, we are working in the perturbative phase, corresponding to the non-perturbative phase of the original theory of fluctuating velocity field.

 \section{Conformal Anomaly and Liouville Field Theory}
 
 Let us scale the powers of $\nu$ out of our variables and parameters.
 After this rescaling, we get a factor of $\nu^{-\twfi}$ in front of the Hamiltonian, which we absorb into the temperature $T_{eff}$
 \begin{equation}
     T_{eff} = \beta^{-1} \nu^{\twfi} \propto  \beta^{-1} \R{}^{-\ot}\ra 0;
 \end{equation}
 
 All parameters in the effective Hamiltonian now are finite in the turbulent limit, and only the effective temperature depends on the viscosity.
 The immediate consequence of this is the appearance of a finite length scale $b$. This parameter is not fixed by minimization because the term $b \vec \nabla^2 \Gamma$ drops from the classical equation for its solution \eqref{solsphere}. 
 
 The kinetic term $H[\Gamma, \vec X ]$ and dissipation term $b E[\Gamma,\vec X]$ have different spatial dependence at distances less and greater than $b$.  It simplifies the equations to introduce the dissipation length
 \begin{equation}
     r_D = 4 b
 \end{equation}
 
\begin{figure}
    \centering
    \includegraphics[width=\textwidth]{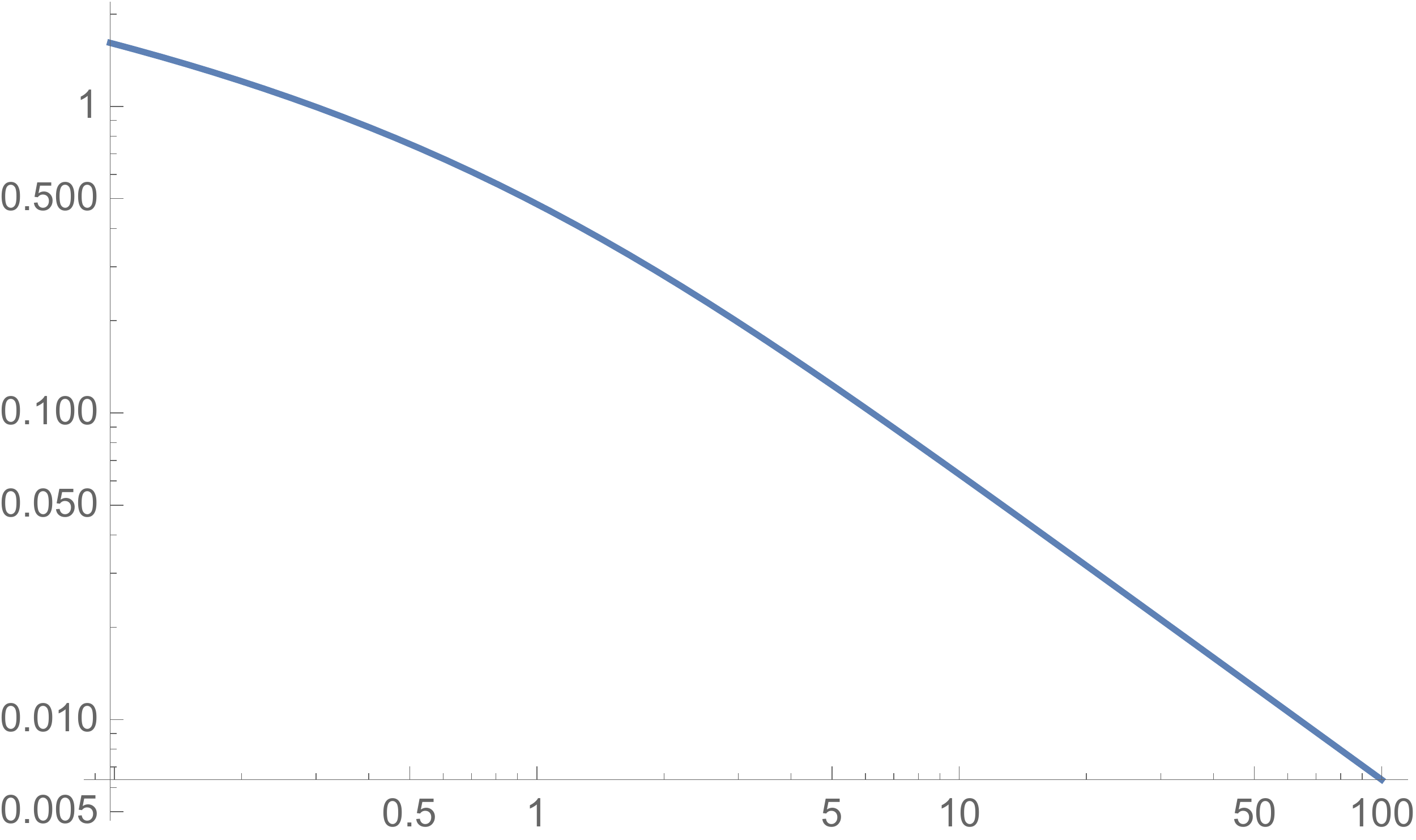}
    \caption{The $\Gamma$ correlator on infinite plane.}
    \label{fig::GammaCorr}
\end{figure}

 Let us study this for a plane (the limit of a large radius of the sphere). In that case, the Hamiltonian for harmonic fluctuations of $\Gamma$  in Fourier space is diagonal
 \begin{eqnarray}
 &&H[\Gamma, \vec X ]+\frac{r_D}{4} E[\Gamma,\vec X] \ra \int \frac{d^2 k}{16\pi^2} |\Gamma_{\vec k}|^2 \left(r_D \vec k^2 + |\vec k|\right)
 \end{eqnarray}
 The corresponding propagator in the coordinate space is related to certain Bessel functions
\begin{eqnarray}
   &&\VVEV{\Gamma(\vec x), \Gamma(0)} = \int \frac{d^2 k e^{\i \vec k \cdot \vec x}}{ \pi^2|\vec k| \left( r_D |\vec k|+ 1\right)} = \frac{1}{r_D} F\left(\frac{|\vec x|}{r_D}\right);\\
   &&F(x) = \pmb{H}_0(x)-Y_0(x);
\end{eqnarray}
The log-log plot of this function is shown in Fig.\ref{fig::GammaCorr}.

 In the UV region of large $\vec k$ we can neglect the kinetic energy term $|\vec k|$ and we are left with a free 2D massless field.
 
 This immediately leads to the conformal anomaly, corresponding to another massless mode, the Polyakov's Liouville field\cite{Pol81} $\varphi$. 
 
 This massless mode comes from the internal metric of the vortex sheet, described by the tangent velocity in the Lagrange equations of motion. Formally, this term in the effective Hamiltonian comes from UV degrees of freedom of the $\delta \Gamma$ fluctuations, and from request that the measure in phase space $\Gamma, \vec X$ stays parametric invariant.
 
 The position field $\vec X$ does not provide the massless excitations. The energy dissipation term $E[\Gamma, \vec X]$ provides a conformally invariant action in the leading expansion in derivatives. The higher order terms, involving the gradient of $\vec X$ times gradients of $\Gamma$ would represent higher dimensional operators of the 2D conformal field theory.
 
The only term with two derivatives of $\vec X$ is the volume term, but it does not lead to kinetic energy for the fluctuations $\delta \vec X$. Therefore, the fluctuations of the  coordinates in the volume do not lead to a conformal action.

 Thus, the effective action (with normalization of \cite{ZZ96} for the Liouville field $\varphi$)
 \begin{align}
    &S = \frac{\beta r_D}{4\Lambda}\int d S(\vec \nabla \Gamma)^2+
    \frac{\beta}{2\Lambda} \int d S\int d S' \frac{\vec \nabla \Gamma \cdot \vec \nabla \Gamma'}{4 \pi |\vec r - \vec r'|} + \\
    &\frac{1}{4 \pi}\int d S\left(  (\vec \nabla \varphi)^2 +4 \Delta e^{2 \varphi} + 4 \hat R \varphi\right);
\end{align}
where $\hat g$ is a background metric, $d S = d^2 \xi \sqrt{\hat g} $ is the area element, $\hat R$ is Gaussian curvature, and $\vec \nabla$ is the gradient in this metric.
The "cosmological constant" $\Delta$ of dimension of $L^{-2}$ comes from the regularization. The natural value for $\mu$ would be $\sim r_D^{-2}$.

This theory without the nonlocal term, describes the inertial range $r \ll r_D$ where there is an unbroken conformal invariance.

Note an important point: while this effective action is quadratic for $\Gamma$, it is essentially nonlinear for the Liouville field.

The normalization of this field follows from a conformal anomaly, and it does not involve temperature. Thus, even in the turbulent limit we have to solve a nonlinear Liouville theory.

\section{Energy Dissipation Correlations}

Fortunately, there is a set of observables which we can compute without knowing the solution of Liouville theory.

We are talking about the correlation functions of the energy dissipation. They can be measured as connected correlators of enstrophy
\begin{eqnarray}
   E(\vec r_1,\dots\vec r_N) = \VVEV{\nu \vec \omega^2(\vec r_1) \dots \nu \vec \omega^2(\vec r_N)}
\end{eqnarray}

\begin{figure}
    \centering
    \includegraphics[width=\textwidth]{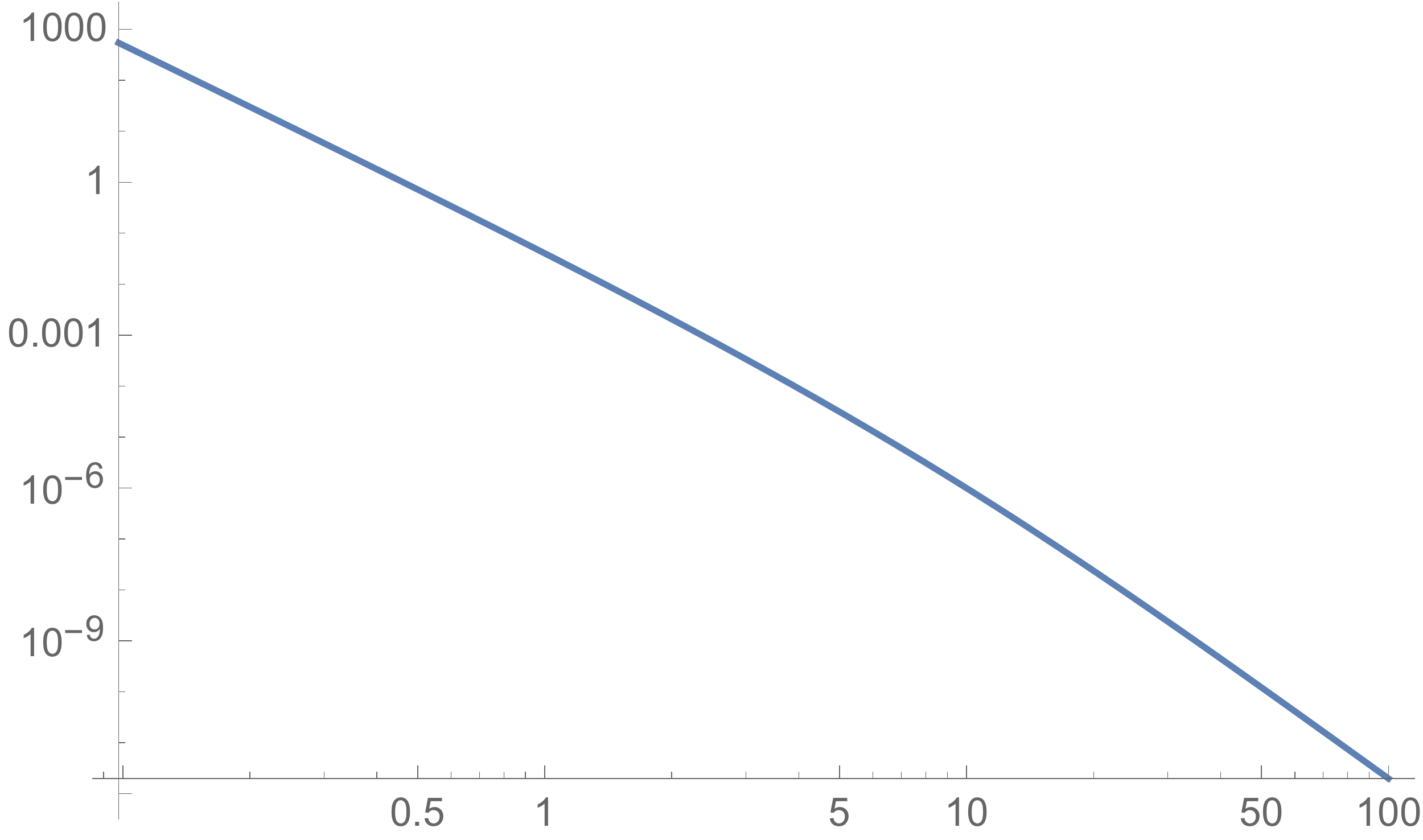}
    \caption{The $(\vec \nabla\Gamma)^2$ correlator on infinite plane.}
    \label{fig::DGDG}
\end{figure}

In our theory, they reduce to the sums of products of the pair correlations of $\vec \nabla \Gamma$ taken in the Gaussian approximation.
For example, the simplest pair correlation
\begin{eqnarray}
   &&E(\vec r_1, \vec r_2) = \Lambda^2\VVEV{(\vec \nabla \Gamma)^2, (\vec \nabla \Gamma)^2} = -\frac{1}{r_D^6} G\left(\frac{|\vec r_1 - \vec r_2|}{r_D}\right);\\
   &&G(x) = F''(x)^2 + \frac{F'(x)^2}{x^2}.
\end{eqnarray}
For reference, we provide the exact expression for these derivatives in terms of Bessel functions:
\begin{align}
    &F''(x) =Y_0(x)-\pmb{H}_0(x)+\frac{\pmb{H}_1(x)-Y_1(x)}{x};\\
    &F'(x) = \frac{1}{\pi }+Y_1(x)+\frac{\pmb{H}_{-1}(x)-\pmb{H}_1(x)}{2}
\end{align}
For small and large arguments, it has power-like asymptotic form
\begin{align}
&G(x) \ra 
\begin{cases}
\frac{8}{\pi ^2 x^4} &\text{for } x \ra 0\text{,}\\
\frac{20}{\pi ^2 x^6} &\text{for } x \ra \infty\text{;}
\end{cases}
\end{align} 
The log-log plot of this function is shown in Fig.\ref{fig::DGDG}.
The shape of this log-log curve is universal, as all scales are eliminated. One can directly match it with DNS or experiments.

Note that this correlation is negative and never changes the sign.

As for the Liouville field, it decoupled from the correlation in the turbulent limit (i.e., low temperature).
It will display itself in thermodynamics.
\section{Thermodynamics}
Once we established (or conjectured) the relation of the Vortex Sheet Turbulence to the Gibbs distribution, we can study its thermodynamics.

Thermodynamics implies changing temperature or some external sources such as pressure and finding changes in the free energy or other thermodynamic potentials. In our case, the temperature is related to the Reynolds number which we can change in real experiments as well as DNS.

\begin{equation}
    T_{eff} \propto \R^{-\ot}
\end{equation}

The specific heat $C_V$ of a Bose system like ours goes to zero as 
\begin{equation}
    C_V \propto T_{eff}^{\frac{n}{2}}
\end{equation}
where $n$ is the number of soft degrees of freedom relevant at small temperatures.

In our case, we have $n=2$, corresponding to $\Gamma$ and $\varphi$.
Therefore, we have a scaling law
\begin{equation}
    C_V \propto T_{eff} \propto \R^{-\ot}
\end{equation}

This estimate only takes into account the kinetic terms in our effective Hamiltonian, thus it applies to phenomena in  the range $ r  \lesssim r_D$.

\section{Circulation Correlations and Liouville Theory}

As we already mentioned\cite{M20c}, this is a rare example of an exactly solvable string theory, namely, Liouville theory, also solved as $c=1$ matrix model of 2D gravity (see the review in \cite{Klebanov91}).

We made some preliminary estimates in \cite{M20c} assuming that the composite field $e^{n\alpha\varphi}$ describes the scaling behavior of $\omega^n$. As we see it now, this estimate is wrong at several levels: because the circulation is not described by this operator, and because of some other peculiarities of the Liouville theory.

Here we are going to take a deeper look at this problem, now that we established that $\Gamma$ rather than $\vec X$ plays the role of the string coordinate.

Let us consider a circular loop $C$ and a Stokes surface inside (a disk). A spherical vorticity sheet $S$ will intersect this disk along some curve $L$ (circular line) as in Fig.\ref{fig::SXC}:
\begin{figure}
    \centering
    \includegraphics[width=\textwidth]{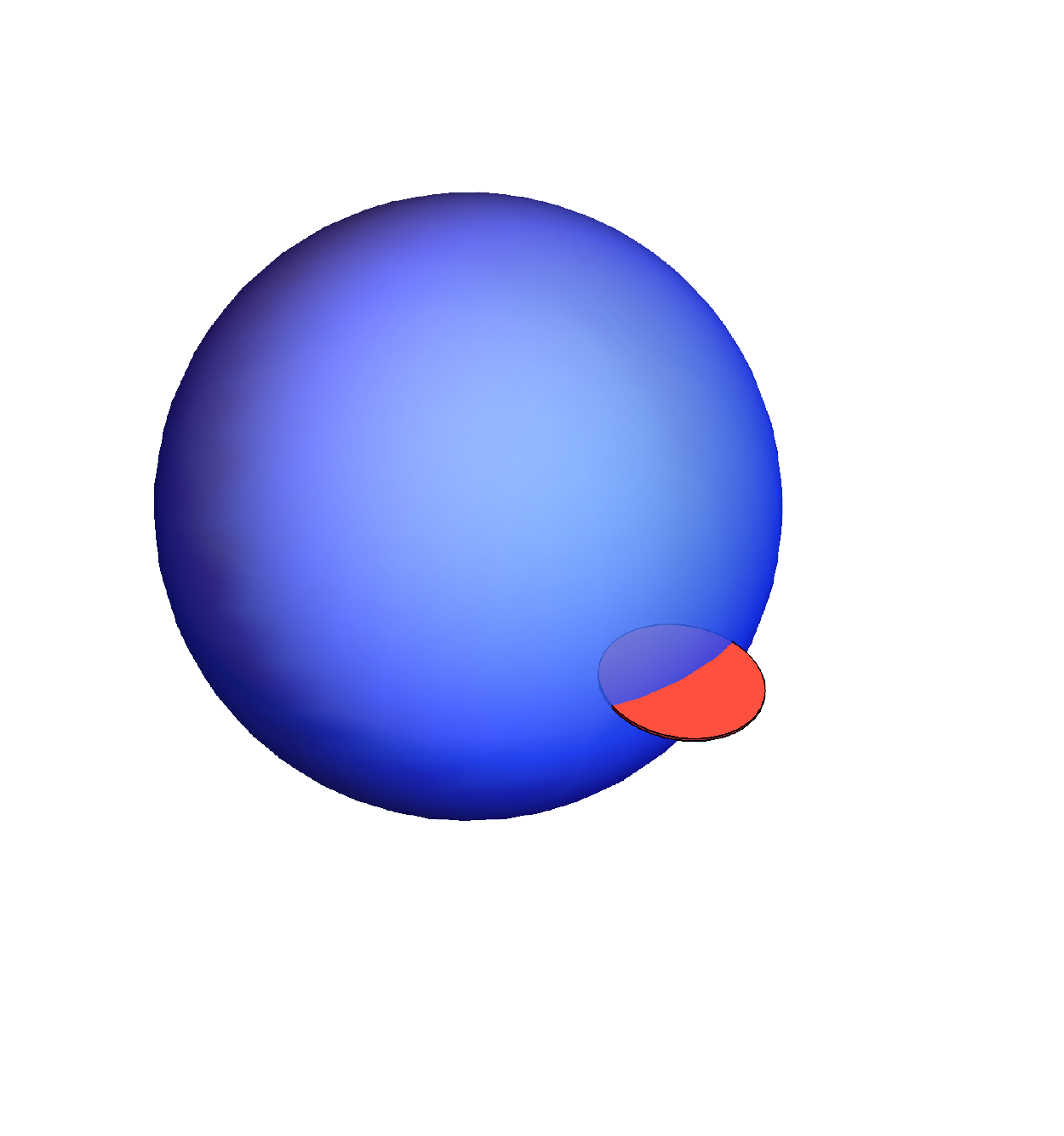}
    \caption{The vortex sheet (blue sphere) intersected by a Stokes surface (red disk)}
    \label{fig::SXC}
\end{figure}
We have the circulation $\Gamma_C$ around the disk reduced to the vortex sheet flux through the line $L$
\begin{eqnarray}
   \Gamma_C = \int_L d \Gamma(\vec r) = \Gamma(\vec b) - \Gamma(\vec a)
\end{eqnarray}
where $\vec a, \vec b \in C$ are endpoints of $L$ at the loop $C$. The metric tensor drops from this relation, so there is no dependence of the Liouville field.

This statement is more general than its proof. Arbitrary closed loop $C$ in 3D space intersects the closed surface $S$ in pairs of points $\vec a_1,\vec b_1,\dots \vec a_N,\vec b_N$, including of course the trivial case $N=0$ (no intersections).

At each point $\vec a_k$ the line $C$ crosses the surface from the outside and then crosses back at $\vec b_k$.

As a consequence, the Wilson loop average reduces to a sum of the terms
\begin{eqnarray}
   &&W_C(g)= \VEV{\EXP{\i g \oint_C d \vec r \cdot \vec v}}=\nonumber\\
   &&\sum_N \VEV{\int d U_N\EXP{\i g \sum_{k=1}^N (\Gamma(\vec b_k) - \Gamma(\vec a_k))} };\\
   && d U_N = d S(\vec a_1)d S(\vec b_1)\dots d S(\vec a_N)d S(\vec b_N)\Pi_C\left(\vec a_1,\vec b_1,\dots,\vec a_N,\vec b_N\right);\\
   &&\Pi_C\left(\vec r_1,\vec r_2,\dots,\vec r_{2N}\right) = \nonumber\\
   &&\prod_{i=1}^{2 N} \int_0^{2\pi} d t_i |\vec C'(t_i)|\delta^3(\vec C(t_i)-\vec r_i)\prod_{i=1}^{2 N} \theta(t_{i+1} - t_i);\\
   && t_{2N+1} \equiv t_1;
\end{eqnarray}

Summing over all surfaces involves the points $\vec a_k, \vec b_k$ sliding along each surface, which is reflected in these integrals $d S$. These points are projected to an ordered set of points $\vec C(t_1),\dots \vec C(t_{2N})$ on a loop, which is reflected in a projection factor $\Pi_C(\dots)$.

 Assuming a low density of the closed surfaces in our ensemble, the $N=1$ term will dominate.
 \begin{subequations}\label{W2}
 \begin{eqnarray}
  && W^{(2)}_C(g) = \VEV{\int d U_2\EXP{\i g  (\Gamma(\vec b) - \Gamma(\vec a))} };\\
  && d U_2 = d S(\vec a)d S(\vec b)\Pi_C\left(\vec a,\vec b\right);\\
   &&\Pi_C\left(\vec r_1,\vec r_{2}\right) = \nonumber\\
   &&\int_0^{2\pi} d t_1 |\vec C'(t_1)|\delta^3(\vec C(t_1)-\vec r_1)\int_{t_1}^{2\pi} d t_2|\vec C'(t_2)|\delta^3(\vec C(t_2)-\vec r_2)
\end{eqnarray}
 \end{subequations}

Now, let us separate the classical part of $\Gamma$ and the fluctuating conformal field $\delta \Gamma$. With the same normalization as the Liouville field
\begin{eqnarray}
   \Gamma = \Gamma^* + \frac{\sqrt{\Lambda}}{\sqrt{\pi \beta r_D}} Y
\end{eqnarray}
With this normalization, the conformal field theory has the form
\begin{eqnarray}
    S_{\mbox{conf}} = 
    \frac{1}{4 \pi}\int d S\left((\vec \nabla Y)^2 +(\vec \nabla \varphi)^2 + 4 \Delta e^{2 \varphi} + 4\hat R \varphi\right);
\end{eqnarray}

The leading low-temperature approximation on a sphere  corresponds to \eqref{solsphere}
\begin{eqnarray}
    &&\Gamma^* = \alpha \vec f \cdot \vec r ;\\
    && \alpha = \frac{2 R^3}{3 \Lambda}
\end{eqnarray}

This provides the classical factor in the loop average, after averaging over the random force $\vec f$
\begin{eqnarray}
    \EXP{-\oh \alpha^2 \sigma \gamma^2 (\vec C(t_1) - \vec C(t_2))^2}
\end{eqnarray}

Integrating this over $\gamma$ and twice over the loop variables $t_1, t_2$ we can estimate the integral at an imaginary saddle point $\gamma^*$, and $t_1, t_2$ near the maximum of distance $D[C] = \max|\vec C(t_1) - \vec C(t_2)|$ between two points on a loop.
\begin{eqnarray}
    && \gamma^* \alpha \sigma  (\vec C(t_1) - \vec C(t_2))^2 = \i \Gamma_C;\\ 
    &&P(\Gamma_C) \sim \EXP{\oh \i \gamma^* \Gamma_C} \sim \EXP{-\oh\frac{\Gamma_C^2}{ \alpha \sigma D[C]^2}}
\end{eqnarray}

Thus, in the classical limit we simply get here a Gaussian distribution variance
\begin{eqnarray}
    \VEV{\Gamma_C^2} \propto D(C)^2
\end{eqnarray}

The power law is the same as we obtained in a previous work, but the exponential tail is missing. We shall discuss this issue in the next section, where we argue that the exponential tail comes from another classical solution-- instanton from \cite{M20c}, or $\mbox{KSL}$  domain wall bounded by an Alice string, as it is known in cosmology and the liquid ${}^3He$. 

We can go one step further and compute the one-loop correction to this classical estimate.
The computation will involve the exponential
\begin{eqnarray}
    &&\VEV{\EXP{ \i q Y(\vec r_1) -\i q Y(\vec r_2)}}_{Liouville};\\
    && q = \rho\gamma;\\
    && \rho = \frac{\sqrt{\Lambda}}{\sqrt{\pi \beta r_D}} 
\end{eqnarray}

The Liouville computations are usually performed for the integrated correlation functions which do not depend upon the points $\vec r_{1,2}$ on a sphere. These functions involve the conformal vertex operators, which we define as
\begin{eqnarray}\label{dressing}
   \mathcal V(q,\vec r) \sim e^{2(1+|q|)\varphi(\vec r) +\i q Y(\vec r)};
\end{eqnarray}
The extra factor $e^{2(1+|q|) \varphi} $ (gravitational dressing) comes from the requirement of conformal invariance. 
This operator has a conformal dimension $\Delta =(1,1)$ which defines its conformal operator product expansion (OPE). 

Strictly speaking, this definition applies only for large negative $\varphi$  when the potential $e^{2\varphi}$ goes to zero so that the theory becomes a free theory.

Whatever happens to this field, at large positive $\varphi$ is a dynamical question to be addressed to experts \cite{ZZ96,Dorn94}. Regardless of this internal mechanics of the Liouville theory, the OPE follows from conformal invariance, and the coefficients of this expansion are calculable using an exact solution of this conformal theory.

For the same reason,--conformal symmetry-- we must use the conformal vertex operator in our correlation function. The conformal invariance is an integral part of the random surface theory, it applies to our Gibbs measure $D\Gamma D\vec X$ in the same way as it applies to the string theory.

The conformal dimension $\Delta = (1,1)$ of the dressed vertex operator does not depend upon the sign in front of $|q|$ in the Liouville term $2(1+|q|) \varphi$.

The positive sign was chosen so that the dressing factor exponentially decays in a free field region $ \phi \ra -\infty$ where the Liouville potential exponentially decays. This is the only choice of the sign which provides a finite and positive two-point function for all real $q$.

The singularity at the origin: $|q| = \sqrt{ \gamma^2}/\rho$ would imply that the odd moments of the PDF for the circulation $\Gamma_C$ do not exist. Moreover, none of the moments exist in case there is a singularity at the origin in Fourier transform.

We do not know at the moment how to resolve this paradox, but we can proceed with the computation. The resulting distribution will be symmetric with finite even moments, within the accuracy of the one-loop computation (corresponding to $q \sim \Lambda^{-\oh} \ra 0$).

In virtue of the conformal invariance, the operator product expansion  $\vec r_1 \ra \vec r_2$ reads
\begin{eqnarray}
    \mathcal V(q,\vec r_1) \mathcal V(-q,\vec r_2) \ra \frac{Z\left(|q|,\Delta\right)}{(\vec r_1 - \vec r_2)^4} 
\end{eqnarray}

The residue $Z(z,\Delta)$ was computed in \cite{ZZ96}. 
We apply their result for the case of fixed area $A= 4\pi R^2$, corresponding to Laplace transform $\tilde Z(|q|, A)$ over the cosmological constant $\Delta$. In their notations, we must set $Q =2, b =1, \alpha = 1 + |q|$.
\begin{eqnarray}
    &&Z(|q|, \Delta) = \int_0^\infty \frac{d A}{A} A^{2|q|}\tilde Z(|q|) \EXP{ - \Delta A} = \Delta^{-2 |q|} \Gamma(2|q|) \tilde Z(|q|);\\
    && \tilde Z(|q|) \propto |q| \frac{\Gamma(1 + 2|q|)}{\Gamma^2(1-2|q|)} = \frac{\sin^2(2\pi q)}{4\pi^2 |q|} \Gamma^3(1 + 2|q|)
\end{eqnarray}
There is an undefined factor $c^{|q|}$ related to the renormalization of the $\Delta$, which we skip here.

We immediately see that $\tilde Z(|q|)$ is positive and does not have any singularities on the real axis in Fourier integral for PDF (see below).
This factor grows as $\EXP{ 6 |q| \log |q|}$ at infinity, but this is beaten by the decrease of the Gaussian factor, so the Fourier integral converges. 

In terms of the QFT analogy, the complete theory (with "gravity" $\varphi$ and "matter" $\Gamma$) has positive decreasing correlation in "target momentum space" $q$, but the pure gravity factor grows as a factorial and is compensated by a Gaussian decrease of the "matter" factor, coming from the vacuum value $\Gamma = \vec q\cdot \vec r$ of the matter.

We have to integrate in \eqref{W2} over the surface which produces a finite dimensionless factor depending on two points $\vec a, \vec b$ on a sphere
\begin{eqnarray}
  &&F_C(\vec a, \vec b) =\int_0^{2\pi} d t_1 |\vec C'(t_1)|\int_{t_1}^{2\pi} d t_2 |\vec C'(t_2)| \int d S(\vec a)\int d S(\vec b)\nonumber\\
  &&\delta^3(\vec C(t_1)-\vec a)\delta^3(\vec C(t_2)-\vec b) =
  \frac{|\vec C'(t_a)||\vec C'(t_b)|}{| \vec C'(t_a) \cdot \Sigma(\vec a)| | \vec C'(t_b) \cdot \Sigma(\vec b)|};\\
  && \vec C(t_a) = \vec a;\\
   && \vec C(t_b) = \vec b;\\
   && \vec \Sigma(\vec r) = \frac{\vec r}{R}
\end{eqnarray}

As a result, we have the following one-loop solution for the loop average
\begin{eqnarray}
    F_C(\vec a, \vec b) \tilde Z\left(\rho|\gamma|,\tilde\Delta\right)\EXP{-\oh \alpha^2 \sigma \gamma^2 (\vec a - \vec b)^2}
\end{eqnarray}

Now, we can define the Fourier integral as twice the integral from zero to infinity of the real part of the Fourier exponential
\begin{eqnarray}\label{PDF}
    &&P(\Gamma_C|\vec a, \vec b) \propto \frac{F_C(\vec a , \vec b)}{|\vec a - \vec b|^4} \Re \int_{0}^\infty d q \frac{\sin^2(2\pi q)}{4\pi^2 q} \Gamma^3(1 + 2q)\nonumber\\
    && \EXP{ \i q \frac{\Gamma_C}{\rho}- \frac{\alpha^2 \sigma}{2 \rho^2} q^2 |\vec a - \vec b|^2};\\
    && \alpha = \frac{2 R^3}{3 \Lambda};\\
    && \rho = \frac{\sqrt{\Lambda}}{\sqrt{\pi \beta r_D}}
\end{eqnarray}

\begin{figure}
    \centering
    \includegraphics[width=\textwidth]{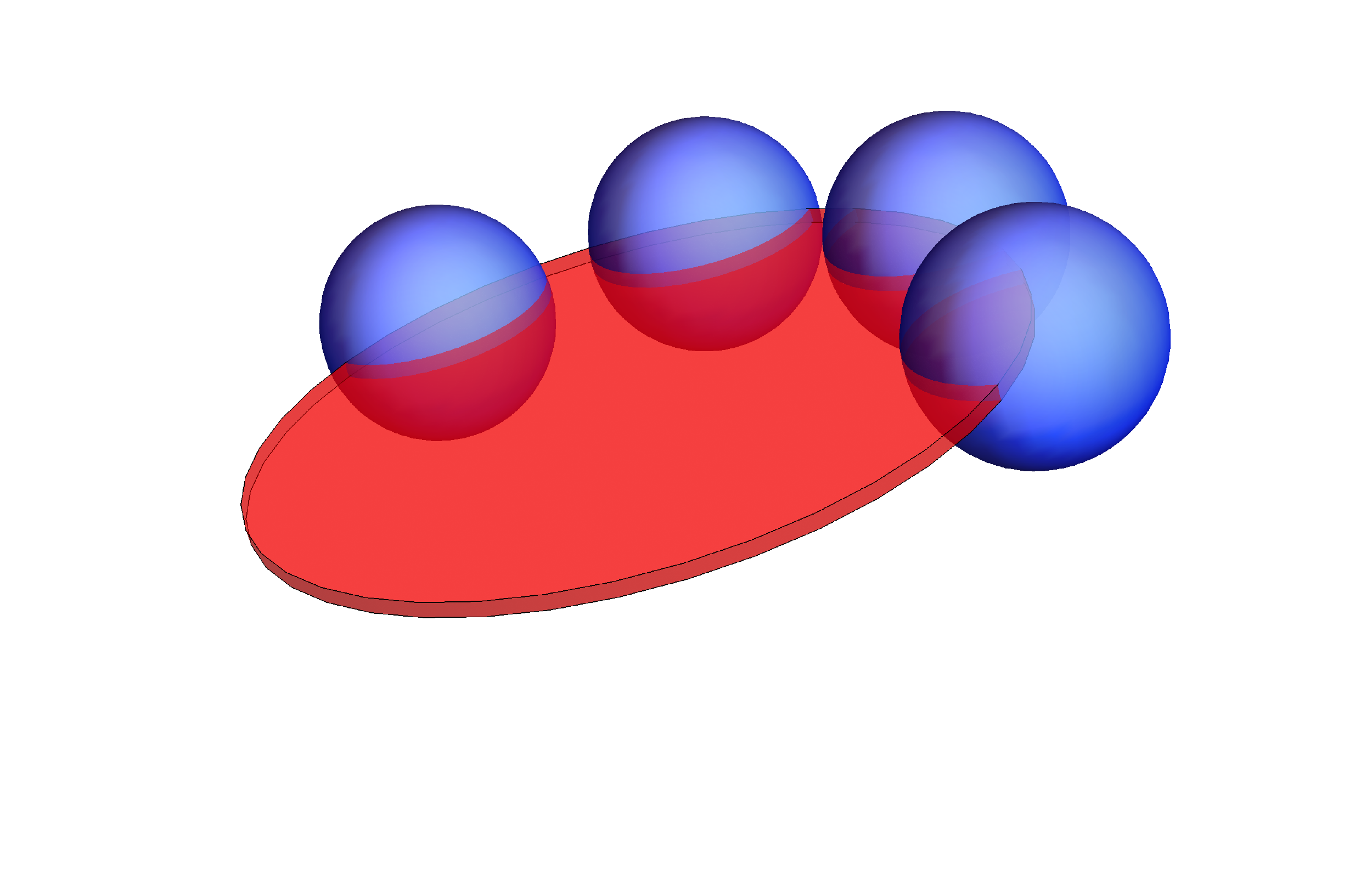}
    \caption{Four spheres intersected by a Disk.}
    \label{fig::Beads}
\end{figure}

In principle, for a finite density of vortex sheets, we should  also add  terms with many surfaces intersected as beads by the same loop.
Fig.\ref{fig::Beads}.

The problem of detecting and summing up the leading terms of low-temperature expansion is well defined in principle, but it deserves a special study in each case like the QED computations, summing the leading logs for scattering amplitudes.

At least we can tell that the Liouville theory provides us with the proper tools to solve the vortex sheet turbulence problem.

One thing is obvious from the representation of this integral \eqref{PDF} for the probability: at a small coordinate difference $D[C] = |\vec a - \vec b|^2 \ra 0$ the integral is dominated by a saddle point at large $q$. 

This saddle point is off the real axis, which leads to oscillations, unacceptable in PDF. However, this is happening beyond the approximation we used to derive this formula.

By our own counting of powers of viscosity, it is not difficult to see that while $\Gamma\sim \Lambda^{-\oh} \ra \infty$, the integration variable is small $q \sim \Lambda \ra 0$. Therefore, the Stirling asymptotic behavior of Gamma function at large $q$ is beyond our one-loop approximation.

In this approximation, we can only expand in q to first order (with $\gamma_E = 0.57721566..$ being the Euler's constant)
\begin{eqnarray}
    &&P(\Gamma_C|\vec a, \vec b) \propto \frac{\alpha \sqrt{\sigma}}{\rho }\frac{F_C(\vec a , \vec b)}{|\vec a - \vec b|^4}   \frac{Re(q_0 -6 \gamma_E q_0^2)}{|\vec a - \vec b|} \EXP{-\oh  \frac{\Gamma_C^2}{\alpha^2 \sigma|\vec a - \vec b|^2}};\\
    && q_0 = \i \frac{\Gamma_C \rho|\vec a - \vec b|^2}{\alpha^2 \sigma }
\end{eqnarray}
We have a preexponential factor
\begin{eqnarray}
   \frac{\alpha \sqrt{\sigma}}{\rho }\frac{6 \gamma_E }{|\vec a - \vec b|}\frac{F_C(\vec a , \vec b)}{|\vec a - \vec b|^4} \frac{\Gamma_C^2 \rho^2|\vec a - \vec b|^4}{\alpha^4 \sigma^2 } = \frac{6 \gamma_E }{|\vec a - \vec b|}\frac{\rho F_C(\vec a , \vec b) \Gamma_C^2}{\alpha^3 \sigma\sqrt{\sigma}}
\end{eqnarray}
This one-loop approximation does not show any time reversal symmetry violation, same as the leading WKB approximation. We are going to study these effects in the next section.

The reader may think -- why bother with this exactly solvable Liouville/matrix gravity if in the end you just used the one-loop approximation.
The answer is -- this is one-loop approximation for our vortex sheet statistics, but a non-perturbative solution for the Liouville theory. Perturbation expansion for the Liouville theory  would correspond to a free Liouville field and it never works. The formulas for conformal dimensions and for the residues in the operator product expansion are non-perturbative. 

Expansion in $q$ is an expansion near the symmetry point of the anomalous dimension $\Delta_q = 1 + q^2$ of the gravitation dressing $e^{2(1+|q|) \varphi}$. This is not the same as the expansion of this exponential in Taylor series, needed for a free field. The exponential potential does not have a well-defined perturbation theory, because it is not bounded. 

Technically, these formulas for $Z(q)$ represent an analytic continuation of the sum of conformal multidimensional integrals which arise in expansion of the exponential $\EXP{ \Delta/\pi\int d S  e^{2 \phi}}$ in powers of the string tension $\Delta$. These conformal integrals were computed in some 2D solvable systems before, so these results were used to compute the sum. 

All we used here in one-loop computation was the Euler constant $\gamma_E$ in the expansion of these gamma functions of the non-perturbatibe solution near the point $q=0$ far away from the perturbation expansion for the Liouville.

 An interesting subject is an extension to surfaces of higher genus. As we already mentioned\cite{M88,AM89,M20c} the variable $\Gamma$ shifts by a period $\Delta_\gamma \Gamma$ when the point goes around some cycle $\gamma$ of a handle on the surface. This means that the mapping of $\Gamma$ on a Riemann surface represents a compactification of our $c=1$ critical string theory. 

This compactification was also discussed in string theory, and there is a solution of the matrix model, summing all topologies. The Liouville theory gives explicit expressions for the correlation on the sphere and the torus, and there are algorithms to compute correlations for higher genus as well. This would be a problem for the future to find any manifestations of higher topology vortex sheets in turbulence.

In higher loop calculations, more results from Liouville theory will be used. This theory and the large $N$ matrix theory providing an alternative solution of this random surface problem, are on the forefront of modern mathematical physics, and we are lucky that the solution was found in the 90-ties and early in this century. Without these theories, our vortex sheet statistics would be just a mathematical abstraction, without any practical use.

\section{KLS domain wall}

We  suggested in a recent work\cite{M20c} a certain topological solution for stationary Euler equations.

It was formulated in terms of Clebsch variables which parametrize the vorticity in terms of a unit vector $\vec S(\vec r) \in S_2$
\begin{equation}\label{CL}
    \vec \omega = Z e_{a b c}  S_a  \vec \nabla S_b \times \vec\nabla S_c;
\end{equation}

where $Z$ is some positive constant, which  becomes a  global variable in thermodynamics.

The solution corresponds to a vorticity sheet like the one above, but with extra normal vorticity at the surface.

We argued that this solution takes part in anomalous dissipation, plus it also explains the exponential tail of the PDF of the velocity circulation.

After extensive discussions with Grigory Volovik, we concluded that my solution is topologically equivalent to the so-called KLS domain wall bounded by Alice string, initially suggested for the early universe.

These stable topological defects\cite{V20,ZH20} were instead observed in real experiments in liquid ${}^3\mbox{He}$, which is an example of a (quantum) fluid with zero viscosity. The dynamic variables in the liquid ${}^3\mbox{He}$ are different from our \CL{} variables, but topology is the same.

The vortex surface is a disk bounded by some loop $C$.

In terms of \CL{} variables, the disk is a domain wall such that $S_3 $ stays continuous, but the the complex field $S_1 + \i S_2$ rotates $2 k+1$ times  by $\pi$ in a complex plane.

In my paper, I assumed that there was an even number of $\pi$ rotations, but only an odd number is topologically stable.

The disk then is a 2D version of the branch cut in the complex plane between two singularities at $\pm a$.

The tangent vorticity in this solution is the same as with ordinary vortex sheets.

In the local tangent frame
\begin{equation}
     \omega_i(x,y,z) = (2 k + 1)\pi Z\delta(z) e_{i j} \d_j (1-S_3(x,y));
\end{equation}

This is the same tangent vorticity we had in vortex sheets with
\begin{equation}
    \Gamma = Z(1-S_3)(2 k +1)\pi
\end{equation} 

The azimuth angle $\varphi = \arg (S_1 + \i S_2)$ has $(2 k +1) \pi$ discontinuity at two sides of the wall.

As it follows from \eqref{CL}, there is also a finite normal component $\omega_z$  of vorticity, also related to the same function $S_3(x,y)$. In cylindrical coordinates $ x = \rho \cos \alpha, y = \rho \sin \alpha$
\begin{equation}
    \omega_z = \frac{1}{(2 k +1)\pi\rho} \pbyp{\Gamma}{\rho}
\end{equation}

\begin{figure}
    \centering
    \includegraphics[width=\textwidth]{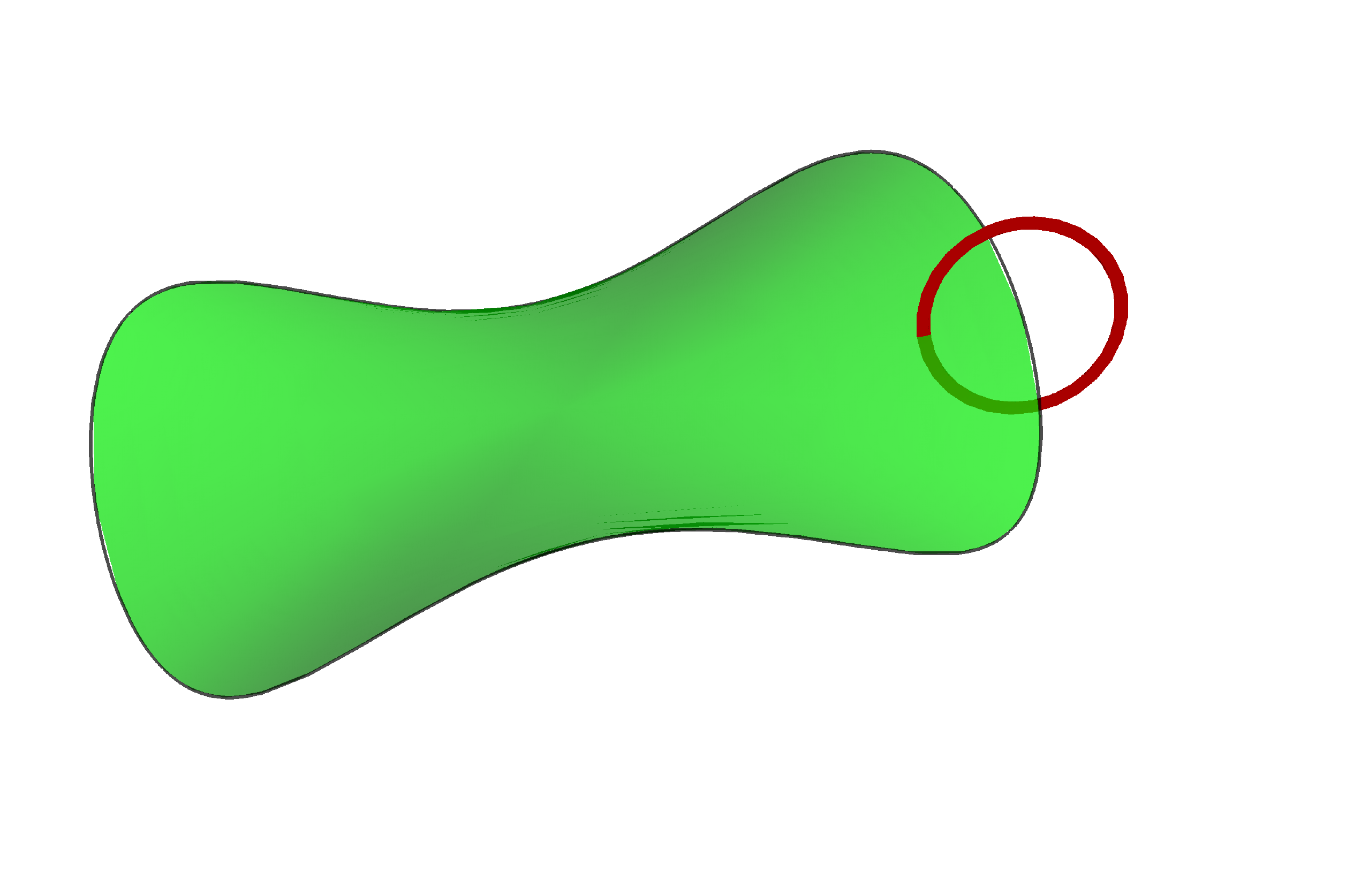}
    \caption{The path (red) around the edge (black) of the KLS domain wall (green).}
    \label{fig::KLS}
\end{figure}

The complex field $\Psi(\vec r) = S_1(\vec r) + \i S_2(\vec r) $ acquires phase $(2k + 1)\pi$ when its coordinate $\vec r$ goes around  the edge of the KLS wall.
(Fig. \ref{fig::KLS})

An example of the $\vec S(\vec r)$ field on a unit disk with this topology would be a stereo mapping from the local tangent plane $x,y$

\begin{align}
    & \rho = x + \i y;\\
    &S_3 = \frac{1 - |\rho|^2}{1 + |\rho|^2};\\
    &\Psi = S_1 + \i S_2 = \frac{2 \rho}{1+ |\rho|^2} \EXP{\i\left(k +\oh\right) \pi \erf{\left(\frac{z}{h\sqrt 2}\right)}};
\end{align}

Here $z$ is a local normal coordinate and $\erf{}$ is an error function, which tends to $\sign(z)$ in the turbulent limit.

The complex field $\Psi$ simply changes sign in that limit, but its gradient $\d_z \Psi $ which enters the tangent vorticity, will come out proportional to $(2k + 1) \pi \delta(z)$. You cannot set $h=0$ before you compute the gradient otherwise you will miss the factor $(2 k + 1)\pi$.

 I assumed that while the normal vorticity is present at the surface, it vanishes outside the viscous layer of the same width $h$ as the tangent component. This will happen if $\vec S(\vec r) \ra \mbox{const}$ away from the disk.
 
There seems to be a contradiction here. The flux of vorticity is supposed to be conserved, so the flux from normal vorticity at $z =0 $ must go somewhere. Maybe there must be a singular vortex line coming out of the disk, say, in its center, and extending to $\pm \infty$?

In other words, the velocity circulation around the edge $C$ of the disk can be computed as a flux of vorticity through arbitrary Stokes surface $S_C$ bounded by $C$.  Choosing this Stokes surface to coincide with the disk itself, we have only the $z$ component of vorticity to contribute to the flux. 

Choosing, say, the semisphere, bounded by $C$ and extended to $z>0$ we would have no contribution from the $z$ component to the flux, but we would have instead the contribution from the $xy$ components of vorticity at the vicinity of the edge. This $\delta(z)$ factor will reduce the semisphere integral to the integral over the loop $C$ with the same result, in agreement with the Stokes theorem.

Thus, we have a tangent vorticity peaking at $z=0$ with Gaussian profile in $z$ direction, plus we have the normal vorticity, finite at $z=0$ and fading out in the normal direction at the viscous width.

The normal component of vorticity contributes to the flux through the disk, but it does not contribute to the velocity field nor  energy dissipation in the turbulent limit $\nu \ra 0, h \ra 0$, because it is present only in a thin layer around the vortex sheet.
 
 The velocity circulation around the loop $C: \rho=R(\alpha)$ reduces to
 \begin{align}
     &\Gamma_C = \int_{S_C} d \vec \sigma 
     \cdot \vec \omega \\
     &=\frac{1}{\left(2 k +1\right)\pi} \oint d \alpha \left(\Gamma(R(\alpha),\alpha) - \Gamma(0,\alpha)\right).
 \end{align}
 
 \section{Energy Balance and Circulation PDF}
 
 The behavior of  $\Gamma(\rho,\alpha)$ is controlled by the minimization of the free energy, leading to the linear integral equation (master equation).
 
 Let us consider the Gibbs distribution for an open KLS domain wall in a thermostat made of closed vortex sheets.

 The boundary $C$ of the open sheet $S_C$ will be a fixed contour in space. The problem is to find the density $\Gamma$ on this sheet.

 As we now understand, this $\Gamma$ must minimize the Hamiltonian of the whole system.

 Assuming known the thermostat velocity $Z\vec v^0$, and $\Gamma= Z \tilde \Gamma$, this problem is equivalent to minimizing the effective Hamiltonian
 \begin{align}
 & H_{eff} = \int d S \int d S' \frac{(\vec \nabla_t \tilde \Gamma) \cdot \vec \nabla'_t \tilde \Gamma') }{8 \pi | \vec r - \vec r'|} +\nonumber\\
 &\mu \Lambda \int d S(\vec\nabla_t  \tilde\Gamma)^2 - \int d S \tilde \Gamma \vec \Sigma \cdot \left(\vec v_0 + \frac{\gamma\vec f}{3}\right)
 \end{align}
 
  Let us now study the energy flow conditions.
  
  The net pumping and dissipation will have the form
  \begin{align}
  &\frac{\mathcal E}{Z} =  \mathcal{P}_T + \frac{1}{3} \vec f \cdot \int_S  \tilde\Gamma^* d \sigma;\\
      &\frac{\mathcal E}{Z^2}= \mathcal{D}_T + \Lambda\int_S (\vec \nabla_t \tilde\Gamma^*)^2;
  \end{align}
  
  where  $\mathcal{P}_T, \mathcal{D}_T$ are contribution of the thermostat and $\tilde\Gamma^*$ is the solution of the master equation.

 Due to the linearity of the master equation, its solution $\tilde \Gamma^*$ linearly depends upon both $\vec f, \vec v_0(.)$, therefore the energy flow would be bilinear
 
 \begin{align}
     &\frac{\mathcal E}{Z} =  \mathcal{P}_T + \vec f\cdot \int_S \hat A(\vec r)\cdot \vec v_0(\vec r) + \vec f\cdot \hat B \cdot \vec f;\\
     &\frac{\mathcal E}{Z^2}= \mathcal{D}_T +  \int_{\vec r \in S}\int_{\vec r' \in S}\vec v_0(\vec r) \cdot \hat C(\vec r,\vec r') \cdot \vec v_0(\vec r')\\
     &+ \vec f \cdot\int_S \hat D(\vec r)\cdot \vec v_0(\vec r) + \vec f\cdot  \hat E \cdot \vec f;
 \end{align}
 
with some matrix kernels $\hat A,\hat C,\hat D$ and some matrices $\hat B, \hat E$.

 Now, the standard thermodynamic procedure is to integrate out the thermostat degrees of freedom.

 In the thermodynamic limit, the thermostat variables in remaining interaction with the subsystem (which is much smaller than the thermostat) are frozen at their equilibrium values (saddle point in the thermostat integral). That includes the global variable $Z$. The expected value of linear terms in $\vec v_0$ would be zero by the space symmetry.

 Solving the above equations for $Z$, we find in the leading order of the small force $\vec f$ 
 \begin{align}
    & Z \ra \frac{\mathcal{P}_T +\vec f\cdot \hat B \cdot \vec f}{\mathcal{D}_T} ;\\
    & \mathcal{E} \ra \frac{\left(\mathcal{P}_T +\vec f\cdot \hat B \cdot \vec f\right)^2}{\mathcal{D}_T}
 \end{align}

 The velocity circulation around the disc will have the form (again, in a leading order in $\vec f$):
 
 \begin{align}
     &\Gamma_C = \frac{Z}{(2 k +1)\pi} \oint d \alpha \left(\tilde \Gamma(R(\alpha),\alpha) - \tilde \Gamma(0,\alpha)\right) \ra\\
     &\frac{\mathcal{P}_T +\vec f\cdot \hat B \cdot \vec f}{(2 k +1)} \Sigma[C];
 \end{align}
 
 Here $\Sigma[C]$ is proportional to the circulation of $\tilde \Gamma $ at $\vec f=0$. It is a linear functional of $\vec v_0$, but this time it is pseudoscalar. Therefore, it is finite. In my paper, I outlined the algorithm for calculating is numerically for a flat loop.

 Naturally, there is always an opposite parity domain wall with $(2 k' +1) = -(2 k + 1)$  and the opposite sign of the circulation (anti-instanton).

We find the circulation proportional to $\mathcal{P}_T +\vec f\cdot \hat B \cdot \vec f$, which leads, after Gaussian integration over random force $\vec f$ with variance $\VEV{f_i f_j} = \sigma \delta_{i j}$, to the algebraic expression for the loop average. Adding the instanton and anti-instanton we find:

\begin{align}
    &\VEV{\EXP{\i \gamma \Gamma_C}} = 
   \oh (W_C(\gamma) + W_C(-\gamma));\\
    & W_C(\gamma) =\frac{\EXP{\i\frac{\gamma \Phi[C]}{2k+1}}}{\sqrt{\det\left(1 - \i\frac{2\gamma \sigma \Sigma[C]}{(2 k +1)} \hat B\right)}};
\end{align}
Let us stress that this is a $3\times 3$ matrix determinant, rather than a functional one.

The pair of root singularities at the imaginary axis leads to the exponential decay:

\begin{align}
    &P_C(\Gamma) = \oh\int_{-\infty}^{\infty} d \gamma  \EXP{-\i \gamma \Gamma} \left( W_C(\gamma) + W_C(-\gamma)\right)\ra\\
    &\mbox{const}\frac{\EXP{- b[C] \left|(2 k +1)\Gamma-\Phi[C]\right|}}{\sqrt{\left|(2 k +1)\Gamma-\Phi[C]\right|} }  + \{\Gamma\Ra -\Gamma\}
\end{align}
This formula with $k =0$ matches very well the DNS by Sreenivasan and Kartik \cite{S19, IBS20}.

One would expect the whole series with higher $k$, providing the smaller exponential correction to the PDF.

\section{Conclusion}

This work's initial goal was to study the steady vortex sheets in Navier-Stokes equations in the turbulent limit of vanishing viscosity at a fixed energy flow and exactly solve these equations in some symmetric cases, which we did. We analyzed at length how various terms in the \NS{} equations add up to zero up to the higher order viscous corrections. 

In doing so, we stumbled upon a very general observation that one can obtain turbulent statistics from the Gibbs distribution corresponding to the vortex sheet dynamics. One should add three extra global constraints:  energy pumping,  energy dissipation,  and the volume inside closed sheets.

The viscosity anomaly leads to an explicit expression for energy dissipation as a kinetic energy for a two-dimensional conformal field $\Gamma$ on a surface. This field would be an Euler integral of motion in the absence of viscosity anomaly.

The energy dissipation is not an integral of motion in the \NS{} equation even in the turbulent limit. We found exact formula for the time derivative of the anomalous dissipation in terms of  $\Gamma$. This turned out to be a cubic functional with four gradients at the surface.

In the low temperature limit when the leading terms represent a conformal theory, this dissipation of dissipation can be discarded as an irrelevant perturbation of the conformal field theory.

The ground manifold (absolute minimum of the effective Hamiltonian for both $\Gamma, \vec X$) corresponds to minimal surfaces, with low-temperature expansion corresponding to small fluctuations around these surfaces as well as fluctuations of $\Gamma$ around the steady solution.

For the closed surface with fixed volume, the minimal shape is a sphere, and for the ones bounded by a steady loop $C$ these are the well-known soap films.

In a turbulent limit, we confirm the scaling laws recently obtained in \cite{M20c} in the context of steady flow.

As a result, we arrive at a particular distribution of discontinuity surfaces in the turbulent statistics. Effective temperature goes to zero in the turbulent limit as $\nu^{\frac{2}{5}} \sim \R^{-\ot}$, at large Reynolds number $\R{}$, so that the low-temperature expansion around the ground manifold would be appropriate. 

The most striking feature of this vortex sheet statistics is that it is equivalent to an exactly solvable critical string theory.
We report here the computation of the simplest correlation functions of this new theory.

We also revised the instanton solution of the previous paper\cite{M20c}, after understanding better its topological meaning and its relation with KLS domain walls in string cosmology, also observed in the quantum superfluid ${}^3He$. The only change was the replacement of an integer winding number in the instanton by a half-integer one, as it is required by topological stability.

The exponential decay as well as the preexponential factor $|\Gamma|^{-\oh}$ stays the same, so that the matching with the DNS data of \cite{S19} still stands.

\section*{Acknowledgments}

I am grateful to  Dmytro Bandak, Gregory Eyink, Nigel Goldenfeld, Igor Klebanov, Sasha Polyakov, Karim Shariff, Katepalli Sreenivasan, Grigory Volovik, Victor Yakhot  and Sasha Zamolodchikov for useful discussions and comments.

This work is supported by a Simons Foundation award ID $686282$ at NYU. 

\appendix
\section{Minimization Problem for a Sphere}

Let us compute the vortex sheet Hamiltonian, momentum and dissipation for the ansatz on a unit sphere $S$
(the powers of the radius follow from dimension counting)
\begin{eqnarray}
    &&\Gamma = \vec q \cdot \vec r;\\
    &&\dal \Gamma = \left(\delta_{\alpha\beta} - \ral\rbe\right) q_\beta.
\end{eqnarray}
We need to compute the following integrals
\begin{eqnarray}
    && H = R^7 q_\alpha q_\beta \int_{\vec r, \vec r' \in S} \frac{ d \ral \wedge d r_\mu d \rbe' \wedge d r'_\mu}{8\pi|\vec r - \vec r'|};\\
    && P_\nu = \frac{R^5}{3} q_\beta \int_{\vec r \in S} d \rbe \wedge d \rga e_{\gamma\mu\nu} r_\mu;\\
    && E = R^2 \Lambda q_\alpha q_\beta \int_{\vec r \in S} \left(\delta_{\alpha\beta} - \ral\rbe\right) 
\end{eqnarray}

The third integral is the simplest. Out of spherical symmetry
\begin{eqnarray}
    \int_{\vec r \in S}\ral\rbe = \frac{4\pi}{3} \delta_{\alpha\beta} 
\end{eqnarray}
In the first and the second integrals we have
\begin{equation}
    d \rbe \wedge d \rga = e_{\beta\gamma\lambda} \rla d S
\end{equation}
where d S is a volume element on $S$ (area element on the unit sphere).
This reduces the momentum integral to the third integral
\begin{eqnarray}
    && P_\nu = \ot R^5 q_\beta e_{\beta\gamma\lambda}e_{\gamma\mu\nu} \int d S r_\mu r_\lambda = \frac{8 \pi}{9} q_\nu R^5
\end{eqnarray}

Finally, in a Hamiltonian, using rotational symmetry once again
\begin{eqnarray}
    &&q_\alpha q_\beta \int_{\vec r, \vec r' \in S} \frac{ d \ral \wedge d r_\mu d r'_\beta \wedge d r'_\mu}{8\pi|\vec r - \vec r'|}=\nonumber\\
    && q_\alpha q_\beta e_{\alpha\mu\rho}e_{\beta\mu\lambda} \int d S \int d S' \frac{r_\rho r'_\lambda}{8\pi|\vec r - \vec r'|}=\nonumber\\
    && \ot q_\alpha q_\beta e_{\alpha\mu\rho}e_{\beta\mu\rho} \int d S \int d S' \frac{\vec r \cdot \vec r'}{8\pi|\vec r - \vec r'|}=\nonumber\\
    && \frac{\vec q^2}{12\pi}  \int d S \int d S' \frac{\vec r \cdot \vec r'}{|\vec r - \vec r'|};
\end{eqnarray}

In the last integral, we shift $\vec r'$ by an infinitesimal amount inside the sphere and use the Legendre polynomial expansion
\begin{eqnarray}
    \frac{1}{|\vec r - \vec r'|} = \sum_0^\infty |\vec r'|^n P_n(\cos \theta)
\end{eqnarray}
Using polar coordinates for $\vec r'$ with North pole at $\vec r$ we have only $P_1(z) = z$ contribute to the expansion
\begin{eqnarray}
    \VEV{\frac{\vec r \cdot \vec r'}{|\vec r - \vec r'|}} = \VEV{\cos^2\theta} = \oh \int_{-1}^{1} z^2 d z = \ot
\end{eqnarray}
This leaves us with 
\begin{eqnarray}
    &&H =R^7 \frac{\vec q^2}{36\pi} (4\pi)^2 = \frac{4\pi\vec q^2 R^7}{9} ;\\
    && \vec P = \frac{8 \pi}{9} \vec q R^5;\\
    && E = \frac{4\pi \Lambda \vec q^2 R^2}{3}
\end{eqnarray}
Minimizing $H - \eta \vec f \cdot \vec P$ with respect to $\vec q$ we find
\begin{eqnarray}
    && \vec q = \frac{\eta \vec f}{R^2};\\
    && \vec f \cdot \vec P = \frac{8 \pi R^3 \vec f^2}{9}\eta;\\
    && E = \frac{4\pi \eta^2\Lambda \vec f^2}{3 R^2}
\end{eqnarray}
Finally, normalizing $\eta,\sigma$ to the mean energy dissipation, we have
\begin{eqnarray}
    &&\mathcal E = \VEV{\vec f \cdot \vec P} = \frac{8 \pi \sigma}{3} \eta R^3;\\
    &&\mathcal E = \VEV{E} = 4\pi \eta^2\Lambda \sigma R^{-2}
\end{eqnarray}
which provides
\begin{eqnarray}
    && \eta = \frac{2 R^5}{3 \Lambda};\\
    &&\sigma = \frac{9 \mathcal E \Lambda}{16 \pi R^8}
\end{eqnarray}

\section{Kolmogorov anomaly}

There is a famous \KO{} anomaly which relates a certain triple correlation function to the same energy flow in an arbitrary space dimension $d$:
\begin{equation}
    \mathcal E  = \lim_{\vec r \ra 0}\pp{r_\beta}\int d^3 r_0\VEV{ \val(\vec r_0) \vbe(\vec r_0) \val(\vec r +\vec r_0)}
\end{equation}
The general triple correlation with two coinciding points can be reconstructed from symmetry, incompressibility, and this relation:
\begin{align}
     &\VEV{\val(\vec r_0) \vbe(\vec r_0) \vga(\vec r + \vec r_0)} =\\
&\frac{\mathcal E }{(d-1)(d+2) V}
	\left(
	 \delta_{\alpha \gamma} r_{\beta} +
	 \delta_{\beta \gamma} r_{\alpha} -
	 \frac{2}{d}\delta_{\alpha \beta} r_{\gamma}
	\right);
\end{align}

The reason for the point splitting in the \KO{} formula is the singularity. 

Formally, at $\vec r =0$  there is a total derivative
\begin{align}
    \val(\vec r_0) \vbe(\vec r_0) \dbe \val(\vec r_0) = \dbe \left(\vbe \frac{\vec v^2}{2}\right) 
\end{align}
so that the integral vanishes in an infinite box with periodic boundary conditions.

Therefore, this relation holds at distances $ \vec r $ larger than the viscous scale.

Let us see how this relation applies to vortex sheets.

The general identity, which follows from the \NS{}if one multiplies both sides by $\vec v$ and averages over an infinite time interval reads:
\begin{align}
     & \int_V d^3 r \VEV{ \nu \vec \omega^2} \\
    &= -\int_V d^3 r \dbe \VEV{\vbe \left(p + \oh \val^2\right) + \nu \val ( \dbe \val - \dal \vbe)}
\end{align}
By the Stokes theorem, the right side reduces to the flow over the boundary $\d V$ of the integration region $V$. The left side is the dissipation in this volume, so we find:
\begin{align}
    &\mathcal E_V = -\int_{\d V} d \vec \sigma \cdot \VEV{\vec v  \left(p + \oh \val^2\right) +\nu \vec \omega \times \vec v }
\end{align}

This identity holds for an arbitrary volume. 
The left side represents the viscous dissipation inside $V$, while the right side represents the energy flow through the boundary $\d V$.

In case there is a finite collection of vortex sheets,  we can expand this volume to an infinite sphere, in which case the $\vec \omega \times \vec v$ term drops as there is no vorticity at infinity.

Furthermore, the velocity in the Biot-Savart law decreases as $|\vec r|^{-3}$ at infinity, so that only the $\vec v p $ term survives
\begin{equation}
    \VEV{\mathcal E_V} \ra -\int_{\d V} d \vec \sigma \cdot \VEV{\vec v  p}
\end{equation}

This energy flow on the right side will stay finite in the limit of the expanding sphere in case the pressure grows as $p \ra - \vec f \cdot \vec r$.
\begin{equation}
    \VEV{\mathcal E} = \vec f_\alpha\lim_{R \ra \infty} R^3\int_{S_2} n_\alpha n_\beta \VEV{ \vbe(R\vec n)}
\end{equation}
This expression is of course, equal to our definition of energy pumping
\begin{align}
    \VEV{\mathcal E} =  \int d^3 r \vec f\cdot\vec v =
    \int_V d^3 r \vec \nabla \left( \vec v (\vec f \cdot \vec r) \right)=
\int_{\d V} d \vec \sigma \cdot \vec v (\vec r \cdot \vec f)
\end{align}

Where did we lose the \KO{} energy flow? It is still there, for any finite volume surrounding the vortex sheet
\begin{align}
   &\VEV{\mathcal E_V} = -\int_V d^3 r \VEV{\vbe \dbe p + \val \vbe\dbe \val} = \\
   & -\int_V d^3 r \VEV{\val \vbe\dbe \val}
  -\int_{\d V} d \vec \sigma \cdot \VEV{\vec v p}
\end{align}
The first term is the \KO{} energy flow inside the volume $V$ and the second one is the energy flow through the boundary.

The pressure can also be expressed in terms of velocity from incompressibility and the \NS{} equation
\begin{align}
    p = - \vec f \cdot \vec r + 
    \int d ^3 \vec r' \frac{\dal \vbe \dbe \val }{ 4\pi | \vec r - \vec r'|}
\end{align}

We see that without finite force $\vec f$ acting on the boundary, say, with periodic boundary conditions, the boundary integral would be absent, and we would recover the \KO{} relation. 

In the conventional approach, based on the time averaging of the \NS{} equations, the periodic Gaussian random force $\vec f(\vec r)$ is added to the right side. In this case, with periodic boundary conditions
\begin{align}
   &\VEV{\mathcal E_V} = -\int_V d^3 r \VEV{\vbe \dbe p  - \vbe f_\beta(\vec r) + \val \vbe\dbe \val} = \\
   & \int_V d^3 r \VEV{\vbe f_\beta(\vec r)}
\end{align}
In the limit when the force becomes uniform in space, we recover our definition as $\mathcal E = \vec f \cdot \vec P$.

Naturally, we assume that the turbulence is a universal phenomenon, so it should not depend upon the mechanism of random forcing nor the boundary conditions. 

As long as there is an energy flow from the boundaries, the confined turbulence in the middle would dissipate this flow in singular vortex structures.

We expect the distribution of these structures to be universal at a given energy flow, regardless how the energy is pumped in.

These assumptions were confirmed in a beautiful experimental work by William Irvine and collaborators in Chicago University (\cite{IM19}). 

They measured the \KO{} energy spectrum, proving that periodic boundary conditions were not necessary.

\bibliography{bibliography}

\end{document}